\documentclass[11pt,letterpaper]{article}

\usepackage{booktabs} 
\usepackage{subcaption}

\usepackage{amsmath}
\usepackage{amsthm}
\usepackage{amssymb}
\usepackage{mathtools}
\usepackage{dsfont}
\usepackage[dvipsnames]{xcolor}
\usepackage{xfrac}
\usepackage{url}
\usepackage{breakurl}

\renewcommand{\paragraph}[1]{{\protect\vspace{8pt}\noindent\sc{#1}}}
\usepackage{multicol}
\usepackage{algorithm}
\usepackage{verbatim}

\usepackage[noend]{algpseudocode}

\usepackage{dsfont}
\usepackage{enumerate}
\newlength{\saveparindent}
\newlength{\saveparskip}
\newcommand{\BE}{\begin{enumerate}} \newcommand{\EE}{\end{enumerate}}
\newcommand{\BI}{\begin{itemize}} \newcommand{\EI}{\end{itemize}}
\newcommand{\BDes}{\begin{description}}\newcommand{\EDes}{\end{description}}
\newtheorem{alg}{Algorithm}
\newcommand{\BA}{\begin{alg}} \newcommand{\EA}{\end{alg}}
 \newtheorem{thm}{Theorem}[section]            
\newcommand{\BT}{\begin{thm}} \newcommand{\ET}{\end{thm}}

\newtheorem{lem}[thm]{Lemma} 
\newcommand{\BL}{\begin{lem}} \newcommand{\EL}{\end{lem}}

\newtheorem{clm}[thm]{Claim}
\newcommand{\BCM}{\begin{clm}} \newcommand{\ECM}{\end{clm}}

\newtheorem{techcor}[thm]{Corollary}
\newcommand{\BCo}{\begin{techcor}} \newcommand{\ECo}{\end{techcor}}

\newtheorem{Conc}[thm]{Conclusion}
\newcommand{\BCONC}{\begin{Conc}} \newcommand{\ECONC}{\end{Conc}}

\newtheorem{Obs}[thm]{Observation}
\newcommand{\BOBS}{\begin{Obs}} \newcommand{\EOBS}{\end{Obs}}

\newtheorem{Exmp}[thm]{Example}
\newcommand{\BEXM}{\begin{Exmp}} \newcommand{\EXMP}{\end{Exmp}}

\newtheorem{cor} [thm] {Corollary}      
\newcommand{\BC}{\begin{cor}} \newcommand{\EC}{\end{cor}}
\newtheorem{prop}[thm]{Proposition}     
\newcommand{\BP}{\begin{prop}} \newcommand {\EP}{\end{prop}}
\newtheorem{conj} {Conjecture}      
\newcommand{\BCJ}{\begin{conj}} \newcommand{\ECJ}{\end{conj}}
\newtheorem{claim} {Claim}      
\newcommand{\BCL}{\begin{claim}} \newcommand{\ECL}{\end{claim}}

\newtheorem{fact}[thm]{Fact}
\newcommand{\BF}{\begin{fact}} \newcommand{\EF}{\end{fact}}
\newtheorem{assumption}[thm]{Assumption}
\newcommand{\BAs}{\begin{assumption}} \newcommand{\EAs}{\end{assumption}}

\newtheorem{defn}{Definition}[section]         
\newcommand{\BD}{\begin{defn}} \newcommand{\ED}{\end{defn}}

\def\FullBox{\hbox{\vrule width 8pt height 8pt depth 0pt}}
\newcommand{\QED}{\;\;\;\FullBox}
\newenvironment{Proof}{\noindent{\bf Proof:~~}}{\hfill\QED}
\newcommand{\BPF}{\begin{Proof}} \newcommand {\EPF}{\end{Proof}}
\newenvironment{proofof}[1]{\noindent{\bf Proof of {#1}:~~}}{\(\hfill\QED\)}
\newcommand{\BPFOF}{\begin{proofof}} \newcommand {\EPFOF}{\end{proofof}}

\newenvironment{smallproof}{\noindent{\bf Proof sketch:~~}}{\(\QED\)}
\newcommand{\bpf}{\begin{smallproof}} \newcommand{\epf}{\end{smallproof}}

\newcommand{\BEQ}{\begin{equation}} \newcommand{\EEQ}{\end{equation}}
\newcommand{\BEQN}{\begin{eqnarray}}\newcommand{\EEQN}{\end{eqnarray}}

\newcommand*{\rom}[1]{\expandafter\@slowromancap\romannumeral #1@}

\newcommand{\argmax}{{\rm argmax}}

\usepackage{mathtools}

\newcommand{\F}{\mathcal{F}}

\newcommand{\SW}[0]{\textsf{SW}}

\newcommand{\vect}[1]{\ensuremath{\mathbf{#1}}}

\newcommand{\goods}{M}

\newcommand{\val}{v}
\newcommand{\vals}{\vect{\val}}

\newcommand{\vali}[1][i]{{\val_{#1}}}

\newcommand{\valtwo}{{\tilde{v}}}
\newcommand{\valtwos}{{\vect{\valtwo}}}
\newcommand{\valthree}{\widehat{v}}
\newcommand{\valfour}{v'}

\newcommand{\price}{p}
\newcommand{\prices}{\vect{\price}}

\newcommand{\alloc}{x}
\newcommand{\allocs}{\vect{\alloc}}

\newcommand{\alloci}[1][i]{{\alloc_{#1}}}

\newcommand{\cutoff}{b}

\allowdisplaybreaks

\begin{document}

\title{Pricing Multi-Unit Markets}

\author{
Tomer Ezra\thanks{Blavatnik School of Computer Science, Tel-Aviv University. Email: \texttt{tomer.ezra@gmail.com}}
\and
Michal Feldman\thanks{Blavatnik School of Computer Science, Tel-Aviv University. Email: \texttt{michal.feldman@cs.tau.ac.il}}
\and
Tim Roughgarden\thanks{Department of Computer Science, Stanford University. Email: \texttt{tim@cs.stanford.edu}}
\and
Warut Suksompong\thanks{Department of Computer Science, Stanford University. Email: \texttt{warut@cs.stanford.edu}}
}

\date{}

\maketitle

\begin{abstract}
  We study the power and limitations of posted prices in multi-unit markets, 
  where agents arrive sequentially in an arbitrary
  order.  We prove upper and lower bounds on the largest fraction of
  the optimal social welfare that can be guaranteed with posted
  prices, under a range of assumptions about the designer's
  information and agents' valuations.
  Our results provide insights about the relative power of uniform and non-uniform prices,
  the relative difficulty of different valuation classes, and the
  implications of different informational assumptions.
  Among other results, we prove constant-factor guarantees for agents with (symmetric) subadditive
  valuations, even in an incomplete-information setting and with
  uniform prices.
\end{abstract}

\section{Introduction}
\label{sec:intro}

We consider the problem of allocating identical items to agents
to maximize the social welfare.  More formally, there are $m$
identical items, each agent $i \in [n]$ has a valuation function $v_i:[m]
\rightarrow \mathbb{R}_{\geq 0}$ describing her value for a given number of
items, and the goal is to compute nonnegative and integral quantities
$q_1,\ldots,q_n$, with $\sum_{i=1}^n q_i \le m$, to maximize the total
value $\sum_{i=1}^n v_i(q_i)$ to the agents.

This problem underlies the design of {\em multi-unit auctions}, which
have played a starring role in the fields of classical and algorithmic
mechanism design, and in both theory and practice.  As with any
welfare-maximization problem, the problem can be solved in principle
using the VCG mechanism.  There has been extensive work on the design
and analysis of more practical multi-unit auctions.  There are
indirect implementations of the VCG mechanism, most famously Ausubel's
ascending {\em clinching auction} for downward-sloping (a.k.a.\
submodular) valuations~\cite{Ausubel04}.  Work in algorithmic
mechanism design has identified mechanisms that retain the
dominant-strategy incentive-compatibility of the VCG mechanism while
running in time polynomial in $n$ and $\log m$ (rather than polynomial
in $n$ and $m$), at the cost of a bounded loss in the social welfare.
Indeed, Nisan \cite{Nisan15} argues that the field of algorithmic
mechanism design can be fruitfully viewed through the lens of
multi-unit auctions.

The multi-unit auction formats used in practice
typically sacrifice dominant-strategy incentive-compatibility in
exchange for simplicity and equitability; a canonical example is the
uniform-price auctions suggested by Milton Friedman (see~\cite{Friedman91})
and used (for example) by the U.S.\ Treasury to sell government
securities.  Uniform-price auctions do not always maximize the social
welfare (e.g., because of demand reduction), but they
do admit good ``price-of-anarchy''
guarantees~\cite{MT15}, meaning that every equilibrium
results in social welfare close to the maximum possible.

A key drawback of all of the mechanisms above is that they require all
agents to participate simultaneously, in order to coordinate their
allocations and respect the supply constraint.  For example, in a
uniform-price auction, all of the agents' bids are used to compute a
market-clearing price-per-unit, which then determines the allocations
of all of the agents.
It is evident from our daily experience that, in many different
markets, buyers arrive and depart asynchronously over time, making
purchasing decisions as a function of their preferences and the
current prices of the goods for sale.\footnote{For examples involving
  identical items, think about general-admission concert tickets,
pizzas at Una Pizza Napoletana (which shuts down for the night when
the dough  runs
out), or shares in an IPO (other than Google~\cite{googleipo}).}
The goal of this paper is to develop theory that explains the efficacy
of such {\em posted prices} in markets where agents arrive
sequentially rather than simultaneously, and that gives guidance on
how to set prices to achieve an approximately welfare-maximizing
outcome.



\subsection{The Model}


We consider a setting where a designer must post prices in advance,
before the arrival of any agents.  We assume that the supply~$m$ is
known.  The designer is given full or incomplete information about
agents' valuations, and must then set a price for each
item.\footnote{No non-trivial guarantees are possible without at least
  partial knowledge about agents' valuations.}
Agents
then arrive in an arbitrary (worst-case) order, with each agent taking
a utility-maximizing bundle (breaking ties arbitrarily), given the
set of items that remain.
These prices are {\em static}, in that they remain fixed
throughout the entire process.
\begin{Exmp}
Suppose $m=3$ and there are two agents, each with the valuation $v(1)
= 5$, $v(2)=9$, and $v(3)=11$, and suppose a designer prices every
item at~4.  The first agent will choose either~1 or~2 items (breaking
the tie arbitrarily).  If the first agent chooses~2 items, the second
agent will take the only item remaining; if the first agent chooses~1
item, then the second agent will take either~1 or~2 items.
\end{Exmp}
In general, we allow different items to receive different prices (as
will be the case in the VCG mechanism for this problem, for example.)
With identical items, however, it is natural to focus on {\em uniform
  prices}, where every item is given the same price.  Generally
speaking, we are most interested in positive results for uniform
prices, and negative results for non-uniform prices.

The overarching goal of this paper is characterize the largest
fraction of the optimal social welfare that can be guaranteed with
posted prices, under a range of assumptions about the designer's
information and agents' valuations.  This goal is inherently
quantitative, but our results also provide qualitative insights, for
example about the relative power of uniform and non-uniform prices,
the relative difficulty of different valuation classes, and the
implications of different informational assumptions.


\subsection{Our Results}

\begin{table}[h]
\begin{center}
   \begin{subtable}{\textwidth}
   \centering
	\begin{tabular}{|c|c|c|c|c|}
		\hline
		 & Uniform prices  & Non-uniform prices \\ \hline
		Submodular & $\frac{1}{2}$ (\ref{t_submod_1_2}, \ref{t_submod_1_2_e}, \ref{t_submod_1_2_identical})& $\frac{2}{3} $ (\ref{t_submod_2_3}, \ref{nonuniformsubmodlowerbound}) [2 items]\\
		
		 &  & $\geq \frac{5}{7}-\frac{1}{m}$ (\ref{t_submod_5_7}),  $\leq 0.802$ (\ref{t_submod_4_5}) [$m$ items] \\   \hline
		
		XOS & $\geq\frac{1}{2}$ (\ref{t_xos_uniform}) & 	$\leq 1-\frac{1}{e}$ (\ref{t_xos_1_e_upper}) \\ \hline
		Subadditive & $\frac{1}{3}$  (\ref{t_subadd_3}, \ref{p_subadd_1_3})
		& $\leq \frac{1}{2}$  (\ref{p_subadd_1_2}) [even with $2$ buyers] \\
		 & $\frac{2}{3}$  (\ref{t_subadd_2_iden}, \ref{p_subadd_2_3_iden}) [$2$ identical buyers] &  $\leq \frac{3}{4}$ (\ref{t_3_4_subadditive_identical}) [even with $2$ identical buyers]  \\
\hline

		General valuations & $\frac{1}{m}$ (\ref{p_general_lower})  & $\frac{1}{m}$ (\ref{t_general_upper})  \\
		\hline
	\end{tabular}
    \caption{Full information}
    \label{t:fullinfo}
    \end{subtable}
   \begin{subtable}{\textwidth}
   \centering
	\begin{tabular}{|c|c|c|c|c|}
		\hline
		 & Uniform prices  & Non-uniform prices \\ \hline
		
		XOS & $\frac{1}{2}$ (\ref{t_xos_uniform}) & 	$\leq 1-\frac{1}{e}$ (\ref{t_xos_1_e_upper}) \\ \hline
		Subadditive & $\geq \frac{1}{4}$  (\ref{t_subadd_uniform})
		& $\leq \frac{1}{2}$  (\ref{p_subadd_1_2}) [even with $2$ buyers] \\
		 &  &  $\leq \frac{3}{4}$ (\ref{t_3_4_subadditive_identical}) [even with $2$ identical buyers]  \\
\hline
	\end{tabular}
    \caption{Incomplete information}
    \label{t:incompleteinfo}
    \end{subtable}
\caption{Summary of results.  All results are new to this paper.  Numbers in parentheses refer to the
  corresponding theorem
or proposition number.}\label{t:sum}
\end{center}
\end{table}

The majority of our results are summarized in Table~\ref{t:sum}; we
highlight a subset of these next.
First, consider the case of a Bayesian setting with  XOS  agent
valuations (see Section~\ref{sec:prelim} for definitions).
That is, each agent's valuation is drawn independently from a known
(possibly agent-specific) distribution over XOS valuations.
Feldman et al. \cite{FGL15} show that, even with non-identical items,
posted prices can always obtain expected welfare at least
$1/2$ times the maximum possible.
This factor of~$1/2$ is tight, even for the special case of a single
item and i.i.d.\ agents.
The posted prices used by Feldman et al. \cite{FGL15} are non-uniform, even when the result
is specialized to the case of identical items
(the price of an item is based on its expected marginal contribution
to an optimal allocation, which can vary across items).
We prove in Theorem~\ref{t_xos_uniform} that with identical items,
and agents with independent (not necessarily identical) XOS
valuations, uniform prices suffice to achieve the best-possible
guarantee of half the optimal expected welfare.
Moreover, this result extends to any class of valuations that is
$c$-close to XOS valuations, with an additional loss of a factor
of~$c$ (Theorem~\ref{t_c_xos}).

While the $1/2$-approximation above is tight for an
incomplete-information setting, this problem is already interesting in
the full-information case where the buyers'
valuations are known (with the order of arrival still worst-case).
Can we improve over the approximation factor of $1/2$ under this
stronger informational assumption?

We prove that uniform prices cannot achieve an approximation factor
better than~$1/2$, even for the more restrictive class of submodular
valuations, and even with two agents
(Proposition~\ref{t_submod_1_2_e}) or identical agents (Proposition~\ref{t_submod_1_2_identical}).
In contrast, with non-uniform prices (still for
submodular valuations),
we prove that an approximation of~$2/3$ is possible
(Theorem~\ref{t_submod_2_3}).
This is tight for the case of two items
(Proposition~\ref{nonuniformsubmodlowerbound}), but in large markets
(with $m \rightarrow \infty$) we show how to obtain an approximation
guarantee of $5/7$ (Theorem~\ref{t_submod_5_7}). In addition, if the order of arrival is known beforehand, we can extract the full optimal welfare (Theorem~\ref{t_submod_opt}).


We next consider the family of subadditive valuations, which strictly
generalize XOS valuations and are regarded as the most challenging
class of valuations that forbid complements.
For example, with non-identical items, it is not known whether or not
posted prices can guarantee a constant fraction of the optimal social
welfare.
For identical items, we prove that this is indeed possible.
In the incomplete-information setting (and identical items), we show
that subadditive valuations are 2-close to XOS valuations
(Proposition~\ref{subadd-xos-bound}), which leads to approximation
factor of $1/4$ (Theorem~\ref{t_subadd_uniform}).
We can also do better in the full-information setting: uniform prices
can guarantee a $1/3$ fraction of the optimal social welfare
(Theorem~\ref{t_subadd_3}), and the approximation is tight (Proposition~\ref{p_subadd_1_3}),
 while even non-uniform prices cannot
guarantee a factor
bigger than $1/2$, even with only two agents (Proposition~\ref{p_subadd_1_2}).
In the case of two identical agents, uniform prices can guarantee a $2/3$ fraction of the optimal welfare (Theorem \ref{t_subadd_2_iden}), and this is tight (Proposition \ref{p_subadd_2_3_iden}).


With all these positive results, the reader might wonder whether constant factor guarantees can be provided for general  valuations.
Unfortunately, this is not the case.
For general valuations, we show that even in the
full-information setting and with non-uniform prices, and even when
there are only two agents and the arrival order is known,
posted prices cannot guarantee more than a $1/m$ fraction
of the optimal social welfare (Proposition~\ref{p_general_lower}).
If the seller can control the
arrival order, however, then even uniform prices can guarantee half of
the optimal social welfare (Theorem~\ref{t_general_2}).
No better bound is possible, even for identical valuations and with
non-uniform prices (Proposition~\ref{p_general_2_upper}).

\subsection{Further Related Work}



The design and analysis of simple mechanisms has been an active area of study in algorithmic mechanism design, particularly within the last decade.  This focus is motivated in part by the observation that simple mechanisms are highly desired in practical scenarios. Examples of simple mechanisms that are used in practice are the generalized second price auctions (GSP) for online advertising \cite{EOS07,Varian07,PaeslemeT10,LucierP11,LucierPT12}, and simultaneous item auctions (where the agents bid separately and simultaneously on multiple items) \cite{CKS08,BhawalkarR12,FFGL13,HKMN11}. These mechanisms are not truthful and are evaluated in equilibrium using the price of anarchy measure.

Posted price mechanisms is perhaps the most prevalent method for selling goods in practice. By simply publishing prices for individual items, these mechanisms are extremely easy to understand and participate in. It should therefore not come as a surprise that posted price mechanisms have been studied extensively for various objective functions (e.g., welfare, revenue, makespan), information structures of values (e.g., full-information, Bayesian, online), and valuation functions (e.g., unit-demand, submodular, XOS). For example, a long line of work has focused on sequential posted prices for revenue maximization and has shown, among other things, that a form of posted price mechanisms can achieve a constant fraction of the optimal revenue for agents with unit-demand valuations \cite{CHK07,CHMS10,CMS10}. Revenue maximization with sequential posted prices has also been studied for a single item, both in large markets \cite{BlumrosenH08} and when the distributions are unknown \cite{BBDS11}, for additive valuations \cite{BILW14,BDHS15}, and for a buyer with complements \cite{EFFTW17}. In several of these works, posted price mechanisms are allowed to discriminate between agents and set different prices for each of them.

In addition to the aforementioned works, a new line of research has considered  dynamic posted prices in online settings such as for the $k$-server and parking problems \cite{CEFJ15}. Moreover, posted price mechanisms have been studied in the context of welfare maximization in matching markets, where prices are dynamic (i.e., can change over the course of the mechanism) but do not depend on the identity of the agents \cite{CEFF16}. Note that the prices used in the mechanisms we consider in this paper are also not discriminatory. Our work, like many others before ours, relies on the Bayesian framework to model a setting with incomplete information; D\"{u}tting et al. \cite{DFKL17} provides a general framework for posted price mechanisms in Bayesian settings.

The sequential arrival of agents considered in the setting of posted price mechanisms fits into the framework of online mechanisms, which deals with dynamic environments with multiple agents having private information \cite{Parkes07}. Our work shows that for identical items and agents with subadditive valuations, posted prices can guarantee a constant fraction of the welfare even while setting the (uniform) prices up front.

\section{Preliminaries}
\label{sec:prelim}

We consider a setting with a set $\goods$ of $m$ \emph{identical} items, and a set $N$ of $n$ buyers.
Each buyer has a valuation function $\vali : 2^\goods \to \mathbb{R}_{\geq 0}$ that indicates his value for
every set of objects. Since items are identical, the valuation depends only on the number of items.
We assume that valuations are monotone non-decreasing (i.e., $\vali(T) \leq \vali(S)$ for $T \subseteq S$) and normalized (i.e., $\vali(\emptyset) = 0$).
We use $\vali(S|T)=\vali(S\cup T)-\vali(T)$ to denote the marginal value of bundle $S$ \textit{given} bundle $T$.

A buyer valuation profile is denoted by $\vals=(\val_1,\dotsc,\val_n)$.
An {\em allocation} is a vector of disjoint sets $\allocs = (\alloc_1, \dotsc, \alloc_n)$, where $\alloci$ denotes the bundle associated with buyer $i \in [n]$ (note that it is not required that all items are allocated). As with valuations, since we consider identical items, an allocation can be represented by the number of items allocated to each buyer.
The {\em social welfare} (\SW) of an allocation $\allocs$ is $\SW(\allocs,\vals)=\sum_{i=1}^{n}\vali(\alloci)$, and the optimal social welfare is denoted by OPT($\vals$).
When clear from the context we omit $\vals$ and write $\SW$ and OPT for the social welfare and optimal social welfare, respectively.

For two valuation functions $\val,\val'$, we say that $\val \geq \val'$ iff $\val(S) \geq \val'(S)$ for every set $S$.
A hierarchy over complement-free valuations is given by Lehmann et al. \cite{LLN06}.
\BD
\label{valfunction}
A valuation function $\val$ is
\begin{itemize}
  \item {\em additive} if $\val(S)= \sum_{i \in S} \val(\{i\})$ for every set $S \subseteq M$.
  \item {\em submodular} if $\val(\{i\}|S) \geq \val(\{i\}|T)$ for every item $i \not\in T$ and sets $S,T$ such that $S \subseteq T\subseteq M$.
  \item {\em XOS} if there exist additive valuation functions $\val^1, \dots , \val^k$ such that $\val(S)=\max_{j=1,\dots,k} \val^j(S)$ for every set $S\subseteq M$.
  \item {\em subadditive} if $\val(S)+\val(T) \geq \val(S \cup T)$ for any sets $S,T \subseteq M$.
\end{itemize}
\ED

Since we assume throughout the paper that all items are identical, we only work with symmetric valuation functions.
	
\BD
A valuation function $\val$ is {\em symmetric} if $\val(S)=\val(T)$ for every sets $S,T\subseteq M$ such that $|S|=|T|$. A symmetric valuation function can thus be represented by a monotone non-decreasing function $\val: \{0,1,\ldots,m\} \to \mathbb{R}_{\geq 0}$, which assigns a non-negative real value to any integer in $[m]$ (recall $\val(0)=0$ as we assume normalized functions).
\ED

In what follows we adjust the definitions of additive, submodular, XOS, and subadditive functions in Definition \ref{valfunction} to the case of symmetric valuation functions. The simplified definition for XOS functions follows from the equivalence between XOS and fractional subadditivity \cite{Feige09}.

\BD
\label{valfunctionsym}
A symmetric valuation function $\val$ is said to be
\begin{itemize}
\item {\em additive} if $\val(i)= a \cdot i$ for every integer $0 \leq i \leq m$ for some constant $a$.
	
  \item {\em submodular} if $\val(i)-\val(i-1)\geq \val(i+1)-\val(i)$ for every integer $1 \leq i \leq m-1$.
  \item {\em XOS} if $\val(i)\geq \frac{i}{j} \cdot\val(j)$ for any integers $1\leq i<j\leq m$.
  \item {\em subadditive} if $\val(i)+\val(j) \geq \val(i+j)$ for any integers $1\leq i,j\leq m$ with $i+j\leq m$.
\end{itemize}
\ED

We assume that the agents arrive sequentially. We will for the most part set static prices for the items, and each arriving agent takes a bundle from the remaining items that maximizes her utility, with ties broken arbitrarily. For some results we will assume dynamic prices, i.e., the seller can set new prices for the remaining items for each iteration (but without knowing which agent will arrive next). If prices $\textbf{p}=(p_1,\dots,p_m)$ are set on the $m$ items, and an agent buys a subset $S$ of them, then her utility is given by $\val(|S|)-\sum_{i\in S}p_i$. For most of the paper we will assume that the arrival order of the agents is unknown, but we will also consider settings where we know this order or where we even have control over the order. We are interested in the social welfare that we can obtain by setting prices in comparison to the optimal social welfare with respect to the worst case arrival order.

Given a buyer valuation profile with submodular valuations $\val_1,\dotsc,\val_n$, we will often consider the multiset $V = \{\vali(j)-\vali(j-1)| i \in [n], j \in [m]\}$, which consists of all marginal values for all buyers and items. Let $\delta$ be the minimum positive difference between any two values in $V$, and let $\epsilon=\delta/2$. Note that the optimal allocation is to sort the marginal values in $V$ in non-increasing order, and allocate items in that order. Thus, OPT is the sum of the $m$ highest marginal values in $V$.

For every value $x$, let $G(x)$ be the number of marginal values in $V$ that are strictly greater than $x$, and let $E(x)$ be the number of marginal values in $V$ that are equal to $x$. Without loss of generality, we will assume that $G(0) \geq m$, since otherwise we can obtain the optimal social welfare by setting all prices to $\epsilon$, which guarantees that all $G(0)$ items yielding positive marginal utility are sold.
Let $\cutoff$ be the unique value satisfying $G(\cutoff) < m$ and $G(\cutoff)+E(\cutoff) \geq m$. Let $m' = G(\cutoff)$; i.e., $m'$ is the number of marginal values strictly greater than $\cutoff$.

\begin{Exmp}
Suppose $m=3$ and there are two agents. The first agent has the valuation $v(1)
= 5$, $v(2)=9$, and $v(3)=11$, while the second agent has the valuation $v(1)=2$, $v(2)=4$, and $v(3)=5$. Then we have $V=\{5,4,2,2,2,1\}$, $\delta=1$, and $\epsilon=0.5$. The optimal welfare is 11 (obtained by either allocating all three items to the first agent, or by allocating two items to the first agent and a single item to the second agent). Moreover, we have $G(2)=2$, $E(2)=3$, $b=2$, and $m'=G(2)=2$.
\end{Exmp}


\section{Properties of symmetric functions}
\label{sec:setfunctions}



In this section, we consider properties of symmetric functions. In addition to being interesting in their own right, these properties will later help us establish welfare guarantees for posted prices (Theorem~\ref{t_subadd_uniform}).

First, we show that every symmetric function admits a unique minimal XOS function as well as a unique minimal submodular function that upper bounds it.

\BP\label{l_xos}
	Let $\val$ be a symmetric function, let $\valtwo$ be the symmetric function $\valtwo(i)=\max_{j\geq i}\frac{i}{j} \cdot \val(j)$. Then $\valtwo$ is XOS, $\valtwo \geq \val$, and for every symmetric XOS function $\valthree$ such that $\valthree\geq \val$, we have $\valthree\geq \valtwo$.	
	\EP

\BPF
	We start by proving that $\valtwo$ is an XOS function.
	By Definition \ref{valfunctionsym} it is enough to show that for every $i,j$ such that
	$i <j$ it holds that $\valtwo(i) \geq  \frac{i}{j} \cdot \valtwo(j)$. Let $i,j$ be such that $i<j$.
	We know that there exists $k\geq j$ such that $\valtwo(j)= \frac{j}{k} \cdot \val(k)$ by the definition of $\valtwo$, and therefore $\valtwo(i) \geq \frac{i}{k} \cdot \val(k) = \frac{i}{k} \cdot \frac{k}{j} \cdot  \valtwo(j)=\frac{i}{j} \cdot \valtwo(j)$, as needed.
	It holds that $\valtwo \geq \val$ since $\valtwo(i) \geq \frac{i}{i}\cdot\val(i)$ for all $i$.

	Now, let $\valthree$ be a symmetric XOS function such that $\valthree\geq \val$, and let $i\leq m$. Let $j=\argmax_{j \geq i} \frac{i}{j} \cdot \val(j)$. We have that $\valthree(j) \geq \val(j)$, and by Definition \ref{valfunctionsym} that $\valthree(i)\geq \frac{i}{j}\cdot \valthree(j) \geq\frac{i}{j}\cdot \val(j) = \valtwo(i)$, as needed.
\EPF

	\BP\label{l_submod}
	Let $\val$ be a symmetric function, and let $\valtwo$ be the symmetric function $\valtwo(i)=\max_{j\geq i, k \leq i}\left(\frac{i-k}{j-k} \cdot (\val(j)- \val(k)) +\val(k)\right)$ (when $j=i=k$ the expression in the maximum is $v(i)$). Then $\valtwo$ is submodular, $\valtwo \geq \val$, and for every symmetric submodular function $\valthree$  such that $\valthree\geq \val$, we have $\valthree\geq \valtwo$.
	
	\EP

\BPF
    We start by proving that $\valtwo$ is a submodular function. To this end, we present an alternative way to define $\valtwo$ that makes its submodularity clear. Plot the points $(i,\val(i))$ for all $i\in[m]$ on a graph, and define $\valfour(0)=0$. If we have defined $\valfour(j)$ for all $j\leq i$, then consider the least $k>i$ such that no point of $\val$ lies strictly above the line connecting $(i,\val(i))$ and $(k,\val(k))$. For all $j$ with $i<j\leq k$, define $\valfour(j)$ so that the point $(j,\valfour(j))$ lies on this line. Since the slope of the lines cannot increase along this process, $\valfour$ is submodular.

    It holds that $\valtwo \geq \val$ since $\valtwo(i) \geq \val(i)$ for all $i$. We show that in fact $\valtwo=\valfour$. Fix any $i\in[m]$, and suppose that $\valfour(i)$ is defined by the line connecting $(r,\val(r))$ and $(s,\val(s))$ for some $r<s$. We have $\valfour(i)=\val(r)+(\val(s)-\val(r))\cdot\frac{i-r}{s-r}$, so $\valfour(i)\leq\valtwo(i)$. Moreover, since all points of $\val$ lie on or below this line, the maximum of $\val(j)+ (\val(k)-\val(j))\cdot \frac{i-j}{k-j}$ is attained at $j=r$ and $k=s$, so $\valfour(i)=\valtwo(i)$.

	Now, let $\valthree$ be any symmetric submodular function such that $\valthree \geq \val$, and let $i\leq m$. Let $r,s=\argmax_{j\leq i, k \geq i}\val(j)+ (\val(k)-\val(j))\cdot \frac{i-j}{k-j}$. We have that $\valthree(r)\geq \val(r)$ and $\valthree(s) \geq \val(s)$.
	Therefore
    \begin{eqnarray*}
			\valthree(i) &\geq &\valthree(r)+ (\valthree(s)-\valthree(r))\cdot \frac{i-r}{s-r} \\
            &= &\valthree(r)\cdot\frac{s-i}{s-r}+\valthree(s)\cdot\frac{i-r}{s-r} \\
            &\geq &\val(r)\cdot\frac{s-i}{s-r}+\val(s)\cdot\frac{i-r}{s-r} \\
			 &= & \val(r)+ (\val(s)-\val(r))\cdot \frac{i-r}{s-r} \\
			 &= &\valtwo(i),
			\end{eqnarray*} as needed.
\EPF

We are interested in approximating functions with ``simpler'' functions. Specifically, for two classes of functions $\mathcal{V}_1\subseteq\mathcal{V}_2$, we want to determine the smallest constant $c$ such that for any function $\val\in\mathcal{V}_2$, there exists a function $\valtwo\in\mathcal{V}_1$ such that $\val\leq\valtwo\leq c\val$. We answer this question for each pair from the classes of subadditive, XOS, and submodular functions and show that the best constant is $c=2$ for all of these pairs. (Note that since all three classes are closed under scalar multiplication, the inequality $\val\leq\valtwo\leq c\val$ above can also be replaced by $\val/c\leq\valtwo\leq \val$.)

		\BP
        \label{subadd-submod-bound}
			For every subadditive function $\val$, there exists a submodular function $\valtwo$ such that $\valtwo \geq \val$ and $\valtwo \leq 2\cdot \val$.
		\EP	

\BPF
			Consider the minimal submodular function $\valtwo$ that upper bounds $\val$ defined in Proposition \ref{l_submod}. For a certain $i$ let $s,r = \argmax_{j\geq i, k \leq i}\frac{i-k}{j-k} \cdot (\val(j)- \val(k)) +\val(k)$.
			It holds that
			\begin{eqnarray*}
			\valtwo(i) &= &\frac{i-r}{s-r}\cdot(\val(s)-\val(r))+\val(r) \\
			 & \leq & \frac{i}{s}\cdot(\val(s)-\val(r))+\val(r) \\
			 &\leq & \frac{i}{s}\cdot(\val(i \cdot \lceil s/i \rceil )-\val(r))+\val(r)\\
			 &  \leq &\frac{i}{s}\cdot(\lceil s/i \rceil \cdot  \val(i)-\val(r))+\val(r)  \\
			 &\leq &\frac{i}{s}\left(\left(\frac{s}{i}+1\right) \cdot  \val(i)-\val(r)\right)+\val(r) \\
			 &  = & \frac{i}{s}\left(\left(\frac{s}{i}+1\right) \cdot  \val(i)\right)+\val(r)\left(1- \frac{i}{s}\right)\\
			 & \leq &  \frac{i}{s}\left(\left(\frac{s}{i}+1\right) \cdot  \val(i)\right)+\val(i)\left(1- \frac{i}{s}\right) \\
			 &= & 2\cdot \val(i),
			\end{eqnarray*}
			where the second and the last inequalities follow from the fact that $\val$ is nondecreasing, and the third inequality follows from the fact that $\val$ is subadditive.
\EPF

Proposition \ref{subadd-submod-bound} immediately yields the following two results.

		\BP
        \label{subadd-xos-bound}
			For every subadditive function $\val$, there exists an XOS function $\valtwo$ such that $\valtwo \geq \val$ and $\valtwo \leq 2\cdot \val$.
		\EP

        \BP
			For every XOS function $\val$, there exists a submodular function $\valtwo$ such that $\valtwo \geq \val$ and $\valtwo \leq 2\cdot \val$.
		\EP

Next, we show that the constant 2 is the best possible for all three pairs of classes. Proposition~\ref{subadd-submod-example} follows directly from Proposition~\ref{xos-submod-example}.

		\BP
        \label{xos-submod-example}
		For every $0\leq c <2$, there exists an XOS function $\val$ such that for every submodular function $\valtwo$ with $\val\leq \valtwo$, we have $\val(i)<\valtwo(i)/c$ for some $i$.
		\EP

\BPF
        Let $l$ be a large integer to be chosen later.
		Consider the XOS function $$
		\val(i)=
		\begin{cases}
		1 & \text{if } i \in \{1,\ldots, \ell\}\\
		\frac{i}{\ell} & \text{if } i \in \{\ell,\ldots,\ell^2\}\\
		\end{cases}.
		$$ By Proposition \ref{l_submod} and setting $j=1,k=\ell^2$, we have that the minimal submodular function $\valtwo$ that upper bounds $\val$ satisfies $$\valtwo(\ell) \geq \val(1) + \frac{\ell-1}{\ell^2-1}\cdot (\val(\ell^2)-\val(1)) = 1+\frac{\ell-1}{\ell+1}. $$ Setting $\ell$ large enough so that $\frac{\ell-1}{\ell+1} > c-1$ yields the desired result.	
\EPF

        \BP
        \label{subadd-submod-example}
		For every $0\leq c <2$, there exists a subadditive function $\val$ such that for every submodular function $\valtwo$ with $\val\leq \valtwo$, we have $\val(i)<\valtwo(i)/c$ for some $i$.
	\EP

        \BP
        \label{subadd-xos-example}
		For every $0\leq c <2$, there exists a subadditive function $\val$ such that for every XOS function $\valtwo$ with $\val\leq \valtwo$, we have $\val(i)<\valtwo(i)/c$ for some $i$.
	\EP

\BPF
Let $l$ be a large integer to be chosen later.
Consider the subadditive function
	$$
	\val(i)=
	\begin{cases}
	1 & \text{if } i \in \{1,\ldots, l\}\\
	2 & \text{if } i = l+1\\
	\end{cases}.
	$$
By Proposition \ref{l_xos} we have that the minimal XOS function $\valtwo$ that upper bounds $\val$ satisfies $\valtwo(l) = \frac{2l}{l+1}$. Setting $l$ high enough so that $\frac{2l}{l+1} > c$ yields the desired result.
\EPF

\section{Submodular valuations}	

In this section we consider submodular valuations.
We first show a reduction from submodular to unit-demand valuations.
In particular we show how to transform $n$ submodular valuations into $nm$ unit-demand valuations, in a way that every static pricing for the $m$ items that guarantees $\rho$ of the optimal social welfare for the unit-demand valuations, will also guarantee $\rho$ of the optimal social welfare for the submodular valuations.
Given this reduction, it is without loss of generality to assume for positive results in Sections \ref{sec:submodnonuniform} and \ref{sec:submoduniform} (which deal with submodular valuations) that all valuations are unit-demand.

Let $\mathcal{M}$ be a market with $m$ items and $n$ agents each with some valuation $v_i:[m]\rightarrow R^+$.
Let $OPT(\mathcal{M})$ be the social welfare of the welfare-maximizing allocation.
In addition, given prices $\prices$, let $SW(\mathcal{M},\prices)$ be the worst case social welfare obtained in market $\mathcal{M}$ under pricing $\prices$.

\BL \label{submod_unitdemand}
Let $\mathcal{M}$ be a market with $n$ agents with symmetric submodular valuations $v_i$'s over $[m]$.
Consider a market $\mathcal{M}'$ with $n\cdot m$ agents, denoted by $(i,j)$ for $i\in[n],j\in[m]$, with symmetric unit-demand valuations $v_i^j$ with values $v_i(j)-v_i(j-1)$. The following holds:
\begin{itemize}
	\item $OPT(\mathcal{M})= OPT(\mathcal{M}')$
	\item For every pricing $\prices$,  $SW(\mathcal{M},\prices)\geq SW(\mathcal{M}',\prices)$.	
\end{itemize}
\EL

\BPF
We prove the two parts in order.
\begin{itemize}
	\item Given an allocation of $\mathcal{M}$ that allocates $x_i$ items to agent $i$.
	The allocation in $\mathcal{M}'$ that allocates items to agents $(i,j)$ if $j \leq x_i$ yields the same social welfare, and therefore $OPT(\mathcal{M}) \leq OPT(\mathcal{M}')$.
	Given an allocation in $\mathcal{M}'$ that allocates items to a set of agents $A=\{(i_1,j_1),\ldots , (i_m,j_m)\}$.
	The allocation in $\mathcal{M}$ that allocates $|\{j: (i,j) \in A \}|$ items to agent $i$ yields at least the same social welfare, therefore  $OPT(\mathcal{M}) \geq OPT(\mathcal{M}')$.
	We get that $OPT(\mathcal{M}) = OPT(\mathcal{M}')$.
	
	\item Let $\sigma$ be the worst case order for market $\mathcal{M}$ with pricing $\prices$, where agent $i$ arrives at turn $\sigma(i)$. And let $x_1,\ldots,x_n$ be the allocation that generates the worst case social welfare.
	Let $\sigma'$ be the order of arrival in market $\mathcal{M}'$ where agent $(i,j)$ arrives at turn $m\cdot(\sigma(i)-1)+j$.
	It is easy to see that the allocation where agent $(i,j)$ is allocated an item if and only if $x_i \geq j$, is a valid allocation with respect to the order of arrival $\sigma'$ and pricing $\prices$. The social welfare of this allocation is  $SW(\mathcal{M},\prices)$, therefore $SW(\mathcal{M},\prices)\geq SW(\mathcal{M}',\prices)$.
\end{itemize}
This completes the proof.
\EPF

\subsection{Non-uniform pricing}\label{sec:submodnonuniform}

We first show that we can obtain $2/3$ of the optimal welfare for submodular valuations if we are allowed to set non-uniform prices.

\BT \label{t_submod_2_3}
For every market with symmetric submodular valuations, there exists a static item pricing $\prices$ that guarantees at least $2/3$ of the optimal social welfare.
\ET

\BPF
By Lemma \ref{submod_unitdemand} it suffices to show the result for buyers with unit-demand valuations. Define $\cutoff,m'$, and $\epsilon$ as in Section \ref{sec:prelim}.
Consider the following pricing structures:

(P1) set a price of $\cutoff-\epsilon$ to all items;

(P2) set a price of $\cutoff-\epsilon$ to $m-m'$ items and a price of $\cutoff+\epsilon$ to $m'$ items.\\
We claim that the maximum of these pricing structures gives at least $2/3$ of OPT.

Under pricing (P1) all $m$ items will be sold, and each one has a marginal value of at least $\cutoff$; thus, a social welfare of at least $m\cutoff$ is guaranteed.


We next show that pricing (P2) guarantees a social welfare of at least $\max\{OPT - (m-m')\cutoff, OPT - m'\cutoff\}$.

The first observation is that every buyer whose value is greater than $b$ will buy an item. This is because when the buyer arrives, at least one item with price $\cutoff+\epsilon$ is available (there are $m'$ such items initially, and the other buyers can buy at most $m'-1$ of them), and an item is worth to her more than $\cutoff+\epsilon$. Therefore, the social welfare of these items is at least $OPT - (m-m')\cutoff$. This also means that the social welfare under pricing (P2) is at least $OPT - (m-m')\cutoff$.

The second observation is that at least $m-2m'$ items that generate value exactly $\cutoff$ will be sold. (It is possible that $m-2m'$ is negative, but this does not affect our analysis.)
This follows from the fact that all $m-m'$ items with price $\cutoff-\epsilon$ will be sold (since these are the cheapest items and the demand is at least $m$), and at most $m'$ of the $m-m'$ items will generate marginal value greater than $\cutoff$ when sold. Therefore, the social welfare from these items is at least $(m-2m')\cutoff$.

Since the sets of items considered in the first and second observations are disjoint, the total social welfare obtained by (P2) is at least their sum, which is $OPT - (m-m')\cutoff + (m-2m')\cutoff = OPT - m'\cutoff$. So the social welfare under pricing (P2) is at least $\max\{OPT - (m-m')\cutoff, OPT - m'\cutoff\}$, as stated.

The average of the guarantee of pricing (P1) and the two guarantees of pricing (P2) is then $(m\cutoff + OPT - (m-m')\cutoff + OPT - m'\cutoff)/3 = 2OPT/3$. Therefore, the best of these two pricing structures gives social welfare at least $2OPT/3$.
\EPF

The following proposition shows that this bound is tight.	
\BP
\label{nonuniformsubmodlowerbound}
There exists a market with two items and two buyers with symmetric submodular valuations such that every static pricing can guarantee a social welfare of at most $2/3$ of the optimal social welfare.
\EP

\BPF
Consider a market with two items and two buyers with the following valuations:
Buyer 1 has a unit-demand valuation with a value of $2$ for each item.
Buyer 2 has an additive valuation with a value of $1$ for each item.
The optimal welfare is $OPT=3$, obtained by allocating to each buyer a single item. We show that no pricing can guarantee more than a welfare of $2$.
Let $\prices=(\price_1,\price_2)$ be some pricing.
Distinguish between the following cases.
\begin{enumerate}
	\item If $p_1,p_2 \leq 1$, then if the additive buyer arrives first, she will buy both items, resulting in a social welfare of 2.
	\item If one of the prices is $\leq 1$ and the other price is $>1$, then if the unit-demand buyer arrives first, she will buy the cheaper item. Then the additive buyer will not buy anything, resulting in a social welfare of 2.
	\item If $p_1,p_2 > 1$, then the additive buyer will not buy any item, thus the social welfare cannot exceed 2.
\end{enumerate}
This concludes the argument.
\EPF

The negative result in Proposition \ref{nonuniformsubmodlowerbound} is obtained for a market with two items.
In what follows we show that the guaranteed social welfare is higher when the number of items is large.

\BT \label{t_submod_5_7}
For every market of $m$ items with symmetric submodular valuations, there exists a static item pricing $\prices$ that guarantees at least $5/7 - 1/m$ of the optimal social welfare.
\ET

\BPF
By Lemma \ref{submod_unitdemand} it suffices to show the result for buyers with unit-demand valuations. Define $V,\cutoff,m'$, and $\epsilon$ as in Section \ref{sec:prelim}.
In addition, let $k = |\{x \in V |x \geq 2\cutoff \}| $. Recall that OPT is the sum of the $m$ highest values in $V$, and therefore $OPT\geq m\cutoff$.

Let $f(i,V)$ be the function that returns the $i$-th highest value in the multiset $V$.
Let $\alpha_1$ be the sum of the highest $\lfloor m'/2 \rfloor$ values in $V$ (i.e., $\alpha_1=\sum_{i\leq \lfloor m'/2 \rfloor}f(i,V)$), and let $\alpha_2$ be the sum of the remaining $\lceil m'/2 \rceil$ values (i.e., $\alpha_2=\sum_{\lfloor m'/2 \rfloor<i\leq m'}f(i,V)$).
Consider the following pricing structures:

(P1) set a price of $\cutoff-\epsilon$ to all items;

(P2) set a price of $\cutoff-\epsilon$ to $m-m'$ items and a price of $\cutoff+\epsilon$ to $m'$ items;

(P3) set a price of $\cutoff-\epsilon$ to $m-k$ items and a price of $2\cutoff-\epsilon$ to $k$ items;

(P4) set a price of $\cutoff-\epsilon$ to $m-\lceil m'/2\rceil$ items and a price of $\cutoff+\epsilon$ to $\lceil m'/2\rceil$ items.\\
We claim that the maximum of these pricing structures gives at least $5/7- 1/m$ of OPT.

Following the same reasoning as in the proof of Theorem~\ref{t_submod_2_3}, pricing (P1) guarantees a social welfare of at least $m\cutoff$ and pricing (P2) guarantees a social welfare of at least $\max\{OPT - (m-m')\cutoff, OPT - m'\cutoff\}$.
We next consider the new pricings (P3) and (P4).

We first show that pricing (P3) guarantees a social welfare of at least $\alpha_1 +(m-m')\cutoff$.
Under pricing (P3) all $k$ items with value at least $2\cutoff$ will be sold, and following the analysis in Theorem~\ref{t_submod_2_3}, at least $m-2k$ additional items will be sold for a value of exactly $\cutoff$.
Thus, the social welfare is at least $\sum_{i \leq k}f(i,V)  + (m-2k)\cutoff$.
We show that this expression is at least $\alpha_1 +(m-m')\cutoff$.
We consider two cases:
\begin{enumerate}
	\item $k < \lfloor m'/2 \rfloor$.
	In this case we have:
	\begin{eqnarray*}
	& & \sum_{i \leq k}f(i,V)  + (m-2k)\cutoff \\
	& = &  \sum_{i \leq  \lfloor m'/2 \rfloor}f(i,V) -\sum_{k<i\leq  \lfloor m'/2 \rfloor}f(i,V) + 2\cutoff(\lfloor m'/2 \rfloor-k)+ (m-2 \lfloor m'/2 \rfloor)\cutoff  \\
	&\geq &  \alpha_1 -\sum_{k<i\leq  \lfloor m'/2 \rfloor}(f(i,V)-2\cutoff) + (m-m')\cutoff \ \ \geq \ \ \alpha_1  + (m-m')\cutoff,  \\
	\end{eqnarray*}
where the last inequality follows from the definition of $k$ (which implies that $f(i,V)<2\cutoff$ for every $i>k$), and the remaining inequalities follow from simple algebraic manipulations.
	
	\item $k \geq \lfloor m'/2 \rfloor$.
	In this case we have:
	\begin{eqnarray*}
		& & \sum_{i \leq k}f(i,V)  + (m-2k)\cutoff \\
		& = &  \sum_{i \leq  \lfloor m'/2 \rfloor}f(i,V) +\sum_{\lfloor m'/2 \rfloor<i \leq k}f(i,V) - 2\cutoff(k-\lfloor m'/2 \rfloor)+ (m-2 \lfloor m'/2 \rfloor)\cutoff  \\
		&\geq &  \alpha_1 + \sum_{\lfloor m'/2 \rfloor<i \leq k}(f(i,V)-2\cutoff) + (m-m')\cutoff \ \ \geq \ \ \alpha_1  + (m-m')\cutoff, 	
	\end{eqnarray*}
	where the last inequality follows from the definition of $k$ (which implies that $f(i,V) \geq 2\cutoff$ for every $i \leq k$), and the remaining inequalities follow from simple algebraic manipulations.
\end{enumerate}

Consider next pricing (P4). Distinguish between two cases:
\begin{enumerate}
	\item
All items are sold. In the worst case the following happens:
\begin{enumerate}
	\item  Items with price $\cutoff+\epsilon$ (there are $\lceil m'/2\rceil$ such items) will generate social welfare from the lowest values in $V$ that are at least $\cutoff+\epsilon$. Thus, the social welfare is at least $\sum_{\lfloor m'/2 \rfloor <i \leq m'}f(i,V) = \alpha_2$.
	\item Items with price $\cutoff-\epsilon$ will generate social welfare of $\cutoff$ each. Thus, the social welfare is at least $(m-\lceil m'/2\rceil) \cutoff$.
\end{enumerate}
Together, if all items are sold, then the social welfare is at least $\alpha_2 + (m-\lceil m'/2\rceil) \cutoff$.
\item There exists an item that remains unsold.
It follows that all values strictly greater than $\cutoff$ contribute to the social welfare (otherwise, there would be no unsold item), yielding a contribution of at least $\alpha_1+\alpha_2$ to the social welfare.
In addition, all items of price $\cutoff-\epsilon$ must be sold.
There are at least $m-m'-\lceil m'/2\rceil$ such items beyond the ones considered toward $\alpha_1+\alpha_2$.
Together, if there exists an unsold item, then the social welfare is at least $\alpha_1 +\alpha_2 + (m-m'- \lceil m'/2\rceil)\cutoff$.
\end{enumerate}
Pricing (P4) guarantees the minimum of the two cases, which is:
$$\min(\alpha_1 +\alpha_2 + (m-m'- \lceil m'/2\rceil)\cutoff,\alpha_2 + (m-\lceil m'/2\rceil)  \cutoff).$$
Note that the first term is chosen from the minimum exactly when $\alpha_1< m'\cutoff$.

We now show that the best of these four pricing schemes guarantees a social welfare of at least $(5/7-1/m)\cdot OPT$.
Assume towards contradiction that all of the above pricing schemes obtain less than $(5/7 -1/m)\cdot OPT$.

\begin{itemize}
\item From pricing (P2) we get that
\begin{eqnarray}
(m-m')\cutoff \geq OPT-\SW(P2) > \frac{2\cdot OPT}{7}. \label{eq_p2} \\
m'\cutoff \geq OPT-\SW(P2) > \frac{2\cdot OPT}{7}. \label{eq_p3}
\end{eqnarray}
\item From pricing (P3) we get that
\begin{eqnarray}
\alpha_2 \geq OPT-\SW(P3) > \frac{2\cdot OPT}{7}. \label{eq_p4}
\end{eqnarray}		
\end{itemize}
We consider two cases:
\begin{enumerate}
	\item $\alpha_1 > m'\cutoff$:
	In this case we get from pricing (P4) that:
	\begin{eqnarray*}
	\alpha_1 -\lfloor m'/2 \rfloor \cutoff & > & OPT-\SW(P4) > \frac{2\cdot OPT}{7} \implies \\
	\alpha_1 &>& \lfloor m'/2 \rfloor \cutoff +\frac{2\cdot OPT}{7} \geq \frac{m'\cutoff}{2} -\frac{\cutoff}{2} + \frac{2\cdot OPT}{7}\\& \stackrel{(\ref{eq_p3})}{>}& \frac{OPT}{7} -\frac{\cutoff}{2}+\frac{2\cdot OPT}{7} = \frac{3\cdot OPT}{7} -\frac{\cutoff}{2}\stackrel{(\ref{eq_p4})}{\implies}  \\
	\alpha_1 +\alpha_2 & \geq &\frac{3\cdot OPT}{7} -\frac{\cutoff}{2} +  \frac{2\cdot OPT}{7} =  \frac{5\cdot OPT}{7} -\frac{\cutoff}{2}.
	\end{eqnarray*}
Since pricing (P2) guarantees a social welfare of $\alpha_1 +\alpha_2$, the last inequality shows that pricing (P2) obtains a social welfare of at least $(5/7 - 1/m)\cdot OPT$ (recall that $\cutoff \leq OPT/m$), a contradiction.

	\item $\alpha_1 \leq m'\cutoff $:
		In this case we get from pricing (P4) that: 		
			\begin{eqnarray*}
			\lceil m'/2 \rceil \cutoff & \geq & OPT-\SW(P4) > \frac{2\cdot OPT}{7} \implies
				m'\cutoff \geq  2 \lceil m'/2 \rceil \cutoff-\cutoff \geq \frac{4 \cdot OPT}{7} -\cutoff \implies  \\
				m\cutoff &= &  (m-m')\cutoff + m'\cutoff \stackrel{(\ref{eq_p2})}{\geq} \frac{2\cdot OPT}{7} -\cutoff +  \frac{4\cdot OPT}{7} =  \frac{6\cdot OPT}{7} -\cutoff.
			\end{eqnarray*}
Since pricing (P1) guarantees a social welfare of at least $m\cutoff$, the last inequality shows that pricing (P1) obtains a social welfare of at least $(5/7 - 1/m)\cdot OPT$ (recall that $\cutoff \leq OPT/m$), a contradiction.
\end{enumerate}
This concludes the argument.
\EPF

The guarantee in Theorem~\ref{t_submod_5_7} approaches $5/7\approx 0.714$ as the number of items grows.
The next theorem shows that this bound cannot exceed 0.802 even for an arbitrarily large number of items.

\BT \label{t_submod_4_5}
For every constant $c$, there exists a market with $m>c$ items with symmetric submodular valuations such that for any static item pricing $\prices$, the social welfare guaranteed by the pricing is at most  $0.802$ of the optimal social welfare.
\ET

\BPF
Consider a market with $m$ items for large enough $m$.
We set $\beta \approx 2.24698$ to be a solution to the equation $x^3-2x^2-x+1=0$, and $\alpha = \frac{1}{(\beta-1)^2}\approx 0.643104$.
An agent is said to be of type 1 (resp., type 2) if she has a unit-demand valuation with value $1$ (resp., $\beta$) for all items.
Consider a market with $m$ type-1 agents and $\lfloor\alpha\cdot m\rfloor$ type-2 agents.
Consider the allocation that gives $\lfloor\alpha\cdot m\rfloor$ items to type-2 agents, and $\lfloor(1-\alpha)\cdot m\rfloor$ items to type-1 agents. This allocation gives a social welfare of $\lfloor\alpha \cdot m\rfloor \cdot \beta + \lfloor (1-\alpha)\cdot m\rfloor \geq (\alpha \cdot m-1) \cdot \beta +  (1-\alpha)\cdot m -1 \geq \frac{\beta}{\beta-1}m-4$.
Given any pricing $\prices$, let $c$ be the number of prices in $\prices$ that are strictly greater than $1$.
We consider two cases:
\begin{itemize}
	\item If $c>\frac{m}{\beta(\beta-1)}=(1-\alpha)m$, then if the type-2 agents arrive first, followed by the type-1 agents, then there will not be any item left for type-1 agents. Therefore, the social welfare is at most $$\alpha \cdot m \cdot \beta  =  m \cdot \frac{ \beta+1}{\beta}
 \leq 0.802 \cdot \left(\frac{\beta}{\beta-1}m-4\right) $$ for large enough $m$.
				
	\item If $c \leq \frac{m}{\beta(\beta-1)}=(1-\alpha)m$, then if the type-1 agents arrive first, followed by the type-2 agents, then the best-case scenario is that the type-1 agents buy the $m-c$ cheap items while the type-2 agents buy the $c$ expensive items. Therefore, the social welfare is at most  $$m-c + c \cdot \beta \leq  m + \frac{m}{\beta} =  m \cdot \frac{ \beta+1}{\beta} \leq 0.802 \cdot \left(\frac{\beta}{\beta-1}m-4\right)$$ for large enough $m$.
\end{itemize}
This concludes the argument.
\EPF

The next result shows that if we know the order of the agents beforehand (while having no control over this order), then we can extract the full optimal welfare.

\BT \label{t_submod_opt}
For every market with symmetric submodular valuations with a known order of arrival, there exists a static pricing $\prices$ that guarantees the optimal social welfare.
\ET

\BPF
Define $\cutoff$ and $\epsilon$ as in Section \ref{sec:prelim}, and number the agents from 1 to $n$ according to their arrival order.
Let $x_1, \ldots , x_n$ be the optimal allocation that allocates the items of marginal value $\cutoff$  greedily (i.e., agent $i$ gets an item of value $\cutoff$ if and only if for all agents $j<i$, agent $j$ also gets all items of marginal value $\cutoff$ that she desires).
Let $k$ be the index of the last agent to get an item of marginal value $\cutoff$.
Consider the pricing $\prices$ where $\sum_{i\leq k} x_i$ items are offered at price $\cutoff-\epsilon$ and $\sum_{i>k} x_i$ items are offered at price $\cutoff+\epsilon$.
This pricing guarantees that agent $i$ will buy exactly $x_i$ items, and therefore guarantees the optimal social welfare as needed.
\EPF

\subsection{Uniform pricing}\label{sec:submoduniform}

Next, we show that if we restrict ourselves to using uniform pricing, we can still guarantee $1/2$ of the optimal welfare. Even though the result also follows from Theorem~\ref{t_xos_uniform}, which holds for XOS valuations in the Bayesian setting, we provide a simpler analysis for this particular case.

\BT
\label{t_submod_1_2}
For every market with symmetric submodular valuations, there exists a static uniform pricing $\price$ that guarantees at least $1/2$ of the optimal social welfare.
\ET

\BPF
As in Theorem \ref{t_submod_2_3}, setting a price of $\cutoff-\epsilon$ to all items guarantees a social welfare of at least $m\cutoff$. On the other hand, if we set a price of $\cutoff+\epsilon$ to all items, then exactly $m'$ items will be sold --- those items with marginal values greater than $\cutoff$ --- and their marginal values will be realized by the buyers. Thus, the social welfare is at least $OPT-(m-m')\cutoff$.

Therefore, using the best of the two prices guarantees at least $\max (m\cutoff,OPT-(m-m')\cutoff)$, which is at least half of $OPT$.
\EPF

The last bound is tight, as shown in the following proposition.

\BP
\label{t_submod_1_2_e}
There exists a market with $m$ items and two buyers with symmetric submodular valuations such that every uniform static pricing yields a social welfare of at most $\frac{m}{2m-1} (\approx \frac{1}{2}) $ of the optimal social welfare.
\EP

\BPF
Consider a market with $m$  items and two buyers with the following valuations:
Buyer 1 has a unit-demand valuation with a value of $m$ for each item.
Buyer 2 has an additive valuation with a value of $1$ for each item.
The optimal welfare is $OPT=2m-1$, obtained by allocating a single item  to the first buyer, and $m-1$ items to the second buyer. We show that no uniform pricing can guarantee more than a welfare of $m$.
Let $\price$ be some uniform pricing.
Distinguish between the following cases.
\begin{enumerate}
	\item If $\price \leq 1$, then if the additive buyer arrives first and buys all items, the social welfare is $m$.
	\item If $\price > 1$, then the additive buyer will never buy any item, and therefore the social welfare is at most $m$.	
	\end{enumerate}
This concludes the argument.
\EPF

The bound $1/2$ for uniform pricing is still tight even with identical buyers.

\BP\label{t_submod_1_2_identical}
There exists a market with identical buyers with symmetric submodular valuation such that every uniform static pricing yields a social welfare of at most $\frac{n+1}{2n} (\approx \frac{1}{2}) $
of the optimal social welfare.
\EP

\BPF
Consider a market with $n^2$  items and $n$ buyers with the following valuation:
$\val(i)=n+i$ for  $i\geq 1$.
The optimal welfare is $OPT=2n^2$, which can be achieved by giving each agent $n$ items. We show that no uniform pricing can guarantee more than a welfare of $n^2+n \approx OPT/2$.
Let $\price$ be some uniform pricing.
Distinguish between the following cases.
\begin{enumerate}
	\item If $\price \leq 1$, then the first buyer will buy all items, and therefore the social welfare is $n^2+n$.
	\item If $\price > 1$, then every buyer will buy at most one item, and therefore the social welfare is at most $n\cdot (n+1)=n^2+n$.	
\end{enumerate}
This concludes the argument.
\EPF

\subsection{Dynamic pricing}

If we allow dynamic pricing, the following result shows that we can extract the full optimal welfare.

\BT \label{t_submod_dynamic}
For every market with $n$ agents with symmetric submodular valuations over $m$ items, there exists a dynamic item pricing that guarantees the optimal social welfare.
\ET

\BPF
We prove the theorem by showing a pricing that maintains the invariant that the current allocation can be completed to an optimal allocation in the induced market.

Define $\cutoff$ and $\epsilon$ as in Section \ref{sec:prelim}. Let $X\subseteq [n], \ell \leq m$ be the remaining available agents and the number of available items in some iteration.
Let $k_i$ be the number of items worth strictly more than $\cutoff$ to agent $i$, and let $y_i$ be the number of items worth exactly $\cutoff$ to agent $i$.
The invariant that the allocation so far can be completed to an optimal allocation implies that
$\sum_{i \in X} k_i \leq \ell$ and $\sum_{i \in X} (k_i +y_i) \geq \ell$.
Consider the following pricing:
If there exists an agent $i \in X$ such that $y_i > \ell-\sum_{j \in X}k_j$, then we set a price of $\cutoff-\epsilon$ to $\min_{i\in X: y_i > \ell-\sum_{j \in X}k_j}k_i+\ell-\sum_{j \in X}k_j$ items, and a price of $\cutoff+\epsilon$ to the remaining items.
Otherwise, we set a price of $\cutoff-\epsilon$ to all items.

We claim that this pricing gives exactly OPT. Fix an iteration, and let $i$ be the arriving buyer in that iteration.
We show that:
\begin{enumerate}
	\item Buyer $i$ buys at least $\max(k_i,\ell   -\sum_{j \in X, j\neq i}(k_j+y_j))$ items.
	\item Buyer $i$ buys at most $\min(\ell-\sum_{j \in X, j\neq i}k_j,k_i+y_i)$ items.
	\item For every number of items $x\in [k_i + \max(0,\ell - \sum_{j\in X}k_j -\sum_{j \in X, j\neq i}y_j), \min(\ell-\sum_{j \in X, j\neq i}k_j,k_i+y_i)]$, it holds that $OPT(X \setminus \{i\},\ell-x) + v_i(x) = OPT(X,\ell)$.
\end{enumerate}
The combination of these properties ensures that the obtained allocation is optimal since in each iteration, the value that agents $i$ gets is equal to the loss of the social welfare for the remaining agents and items.

We now prove the three aforementioned properties.

\begin{enumerate}
	\item
	Firstly, if there exists $i'\in X$ such that $i' \neq i$ and $y_{i'} > \ell-\sum_{j \in X}k_j$, then $\ell  -\sum_{j \in X, j\neq i}(k_j + y_j) \leq k_i +\ell - \sum_{j\in X}k_j -y_{i'}<k_i$, and therefore it is enough to show that agent $i$ will buy at least $k_i$ items.  Agent $i$ buys at least $k_i$ items  since $\ell \geq \sum_{j \in X} k_j \geq k_i$ and therefore at least $k_i$ items are available at price no more than $\cutoff + \epsilon$.

	Secondly, if the only agent $i'\in X$ such that $y_{i'} > \ell-\sum_{j \in X}k_j$ is agent $i$, then we have
	$\min_{i'\in X: y_{i'} > \ell-\sum_{j \in X}k_j}k_{i'} = k_i$, and we get that the number of items that are offered at price $\cutoff-\epsilon$ is  $k_i+\ell-\sum_{j \in X}k_j \leq k_i+y_i$.
	Therefore agent $i$ will buy exactly $k_i+\ell-\sum_{j \in X}k_j\geq  \max(k_i,\ell -\sum_{j \in X, j\neq i}(k_j+y_j))$ items, since $\ell - \sum_{j\in X}k_j \geq 0$.

	Otherwise, there is no agent $i'$ such that $y_{i'} > \ell-\sum_{j \in X}k_j$, and all items are priced at $\cutoff-\epsilon$.
	Therefore agent $i$ buys $k_i+y_i \geq \ell  -\sum_{j \in X, j\neq i}(k_j + y_j)$ items, since $ \sum_{j\in X}(k_j+y_j) \geq \ell$.
	
	\item If $y_i > \ell-\sum_{j \in X}k_j$, then $\min_{i'\in X: y_{i'} > \ell-\sum_{j \in X}k_j}k_{i'} \leq k_i $, and the number of items at price $\cutoff-\epsilon$ is at most $\ell-\sum_{j\in X, j\neq i}k_j$. Since $k_i\leq \ell-\sum_{j\in X, j\neq i}k_j$, agent $i$ buys at most $\ell-\sum_{j\in X, j\neq i}k_j$ items.
	Otherwise, agent $i$ buys at most $k_i+y_i \leq k_i +\ell-\sum_{j \in X}k_j =\ell-\sum_{j\in X, j\neq i}k_j$ items.
	
    When all prices are at least $\cutoff -\epsilon$, agent $i$ does not buy more than $k_i + y_i$ items.	
	\item We have
	\begin{eqnarray*}
		OPT(X,\ell) &=& \sum_{j \in X} v_j(k_j) + \left(\ell -\sum_{j\in X}k_j\right)\cdot\cutoff \\
		& = &  v_i(x) + \sum_{j \in X, j\neq i} v_j(k_j) +\left(\ell - x -\sum_{j \in X, j\neq i}k_j\right)\cdot\cutoff \\
		& = &v_i(x) + OPT(X\backslash\{i\},\ell-x).
	\end{eqnarray*}
	The second equation follows from the fact that $k_i \leq x \leq k_i + y_i$, implying that $v_i(x)=v_i(k_i) + (x-k_i)\cutoff$.
	For the last equation, we need to show that (a) $\ell - x -\sum_{j \in X, j\neq i}k_j \geq 0$ (i.e., there are enough items), and (b) $\ell - x -\sum_{j \in X, j\neq i}k_j \leq \sum_{j \in X, j\neq i}y_j$ (i.e., all items will be bought). Part (a) follows directly from (2) and part (b) follows directly from (1).
\end{enumerate}
Therefore all three properties hold, and the proof is complete.
\EPF

\section{XOS valuations}

In this section we consider XOS valuations. We give upper bounds on the approximation ratio for both static and dynamic pricing.

\BT \label{t_xos_1_e_upper}
There exists a market of $m$ items and two agents with symmetric XOS valuations for which no static pricing yields more the $1-1/e$ of the optimal social welfare.
\ET

\BPF
Consider a market with $m$ items for large enough $m$.
Let $k=\lfloor m/e \rfloor$.
Assume the following agents' valuations:
$$
\val_1(i)=
\begin{cases}
k & i < k;\\
i & i \geq k;\\
\end{cases}
$$
$$
\val_2(i)=
\begin{cases}
\sum_{0 \leq j < m-k} \frac{m-k-j}{m-1-j} & i > m- k;\\
\sum_{0 \leq j < i} \frac{m-k-j}{m-1-j} & i  \leq m-k.\\
\end{cases}
$$
One can check that $v_1$ is XOS and $v_2$ is submodular.
The allocation that maximizes the social welfare is the one that gives all items to agent $1$, which generates a social welfare of $m$.
We show that if agent $2$ arrives first, then for every pricing $\prices$, agent $1$ will not buy more than one item.
Assume agent $2$ bought $s$ items, so that all remaining items have a price of at least $\frac{m-k-s}{m-1-s}$ (otherwise agent $2$ would have bought more items).
Agent $1$'s marginal utility from buying $i$ items beyond a single item is:

$$
u_1(i+1) - u_1(1) \leq
\begin{cases}
-i\cdot \frac{m-k-s}{m-1-s} & i < k;\\
i-k - i\cdot \frac{m-k-s}{m-1-s}  & m-s > i \geq k.\\
\end{cases} 
$$
For the case where $i<k$, the marginal utility is clearly smaller than or equal to $0$.
For the case where $m-s > i \geq k$, the maximum marginal utility is obtained for $i=m-s-1$, where it is also smaller than or equal to $0$.
We conclude that agent $1$ buys at most a single item, if agent $2$ arrives first.
Therefore the social welfare is bounded by  $\val_1(1) + \val_2(m) = k + \sum_{0 \leq j < m-k} \frac{m-k-j}{m-1-j} = m - \sum_{0 \leq j < m-k} \frac{k-1}{m-1-j}= m - \sum_{k \leq j < m} \frac{k-1}{j} \approx m -k+1 \approx (1-1/e)\cdot OPT,$
where the last two equalities follow by the choice of $k=\lfloor m/e \rfloor$.
\EPF

\BT \label{t_xos_dynamic_5_6_upper}
There exists a market of three items and two agents with symmetric XOS valuations for which no dynamic pricing yields more the $5/6$ of the optimal social welfare.
\ET

\BPF
Consider a market with two agents and three items, with the following valuations:
$$
\val_1(i)=
\begin{cases}
4 & i \leq 2;\\
6 & i = 3;\\
\end{cases}
$$
and the second agent has a unit-demand valuation with value 1.
The allocation that maximizes the social welfare is the one that gives all items to agent $1$, generating a social welfare of $6$.
Let $\prices$ be the prices in the first iteration.
If there exists an item $j$ with price $\price_j \leq 1$, then if agent $2$ arrives first, she will buy item $j$, and the social welfare will be at most $5$.
If all items have prices greater than $1$, then if agent $1$ arrives first, then she will buy at most one item, and again the social welfare will be at most $5$.
\EPF

\section{Subadditive valuations}

In this section we consider subadditive valuations. Our main result of this section is the existence of a uniform price that guarantees at least $1/3$ of the optimal welfare.

\subsection{Upper bounds}

\BT \label{t_subadd_3}
For every market of $m$ items with symmetric subadditive valuations, there exists a uniform static item pricing $\prices$ that guarantees at least $1/3$ of the optimal social welfare.
\ET

\BPF
We begin by establishing some technical lemmas. The first lemma provides a guarantee on the welfare obtained by an agent at a certain price in terms of the number of items left when the agent arrives.

\BL \label{l_kv2}
Let $\val$ be a symmetric subadditive valuation function, and let $\valtwo$ be the minimal submodular function such that $\valtwo \geq \val$, as defined in Proposition~\ref{l_submod}.
Let $\cutoff_k = \valtwo(k)-\valtwo(k-1)$ be the $k$-th marginal value of $\valtwo$.
When an agent with valuation $\val$ arrives, if there are $k$ available items, all priced at $\price=\cutoff_k/2$, then the welfare obtained from this agent is at least $k\cutoff_k/2$.
\EL

\BPF
For notational simplicity we write $\cutoff = \cutoff_k$.
Let $\ell=\max_i \{\valtwo(i)-\valtwo(i-1)\geq \cutoff \}$; notice that $ \ell \geq k $ by definition. Let $\alpha$ be the integer such that $\frac{\ell}{\alpha} \leq k < \frac{\ell}{\alpha-1}$.
We can assume without loss of generality that $\alpha \geq 2$, since for $\alpha =1$ the agent will buy $k~ (=\ell)$ items and $\val(k)=\valtwo(k)\geq k\cutoff$/2).
Let $c=\max_i \{\valtwo(i)-\valtwo(i-1) > \cutoff \}$; notice that $ c < k $ by submodularity of $\valtwo$.

Let $u(x)=\val(x)-x\cdot \cutoff/2$ be the utility that an agent with valuation $\val$ derives from obtaining $x$ items.
Let $\underline{a} = \left\lfloor\frac{\ell}{\alpha}\right\rfloor $, $\overline{a} = \left\lceil\frac{\ell}{\alpha}\right\rceil $,  and $\beta\equiv\ell\pmod{\alpha}$ with $0\leq\beta<\alpha$.
By simple algebra it holds that
\begin{equation}
\label{eq:mod}
(\alpha -\beta)\cdot \underline{a} + \beta \cdot \overline{a} = \ell.
\end{equation}

We have:
\begin{align*}
(\alpha -\beta)\cdot u(\underline{a}) + \beta \cdot u(\overline{a})
&=(\alpha -\beta)\cdot \left(\val(\underline{a})-\underline{a}\cdot \frac{\cutoff}{2}\right) + \beta \cdot \left(\val(\overline{a}) - \overline{a} \cdot \frac{\cutoff}{2}\right)\\
& = (\alpha -\beta)\cdot \val(\underline{a}) + \beta \cdot \val(\overline{a})  - \beta \cdot \overline{a} \cdot \frac{\cutoff}{2}-(\alpha -\beta)\cdot\underline{a}\cdot \frac{\cutoff}{2}
\geq
\val(\ell) - \frac{\ell \cutoff}{2},
\end{align*}
where the last inequality follows from the subadditivity of $\val$ and Equation~(\ref{eq:mod}).
Since the first expression is a sum of $\alpha$ terms, there exists a number of items $x'\in\{\underline{a},\overline{a}\}$ such that $u(x') \geq \frac{\val(\ell) - \frac{\ell \cutoff}{2}}{\alpha}$. Now, let $x^*$ be the number of items that the agent actually buys. We know that $x^*\geq c$ since the first $c$ items yield an average marginal value of more than $b$. Moreover, since the agent buys $x^*$ items when she has a choice of buying $x'$, we have $u(x^*)\geq u(x')\geq \frac{\val(\ell) - \frac{\ell \cutoff}{2}}{\alpha}$.
We now consider two cases:

\begin{enumerate}
\item $\val(c) \geq k\cutoff/2$.
 In this case since we have that $x^*\geq c$ it follows immediately that $\val(x^*) \geq \val(c) \geq k\cutoff/2$ as needed.
	

\item $\val(c) < k\cutoff/2$.
Since all marginal values beyond the $c$th item are no greater than $\cutoff$, we have $ \val(c+x) \leq  \val(c) +\cutoff x$ for any $x\geq 0$. This implies that $u(c+x) \leq u(c) +x\cutoff/2 $ .
In particular, since $x^* \geq c$ and $x^*$ yields the optimal utility to the agent, we have $u(c)\leq u(x^*)\leq u(c)+(x^*-c)b/2$.
This implies that $x^* \geq  c + \left\lceil \frac{u(x^*) -u(c)}{\cutoff/2} \right\rceil $,
and it follows that:
\begin{align*}
\val(x^*)  & = \val(c) + (x^*-c) \cdot \frac{\cutoff}{2} + u(x^*) -u(c)
\geq \val(c) + \left\lceil \frac{u(x^*) -u(c)}{\cutoff/2} \right\rceil \cdot \frac{\cutoff}{2} + u(x^*) -u(c) \\
& \geq \val(c) + u(x^*) -u(c)  + u(x^*) -u(c)
= \val(c) +2u(x^*) -2u(c) \\
& \geq \val(c) + 2\cdot\frac{\val(\ell) - \frac{\ell \cutoff}{2}}{\alpha} -2\left(\val(c)- \frac{c\cutoff}{2}\right)
= \frac{2\val(\ell) -\ell \cutoff - \alpha \val(c) +\alpha c\cutoff}{\alpha} \\
& = \frac{2(\val(c) + (\ell-c)\cutoff) -\ell \cutoff - \alpha \val(c) +\alpha c\cutoff}{\alpha}
= \frac{\val(c)(2-\alpha) + c\cutoff (\alpha-2) +\ell \cutoff }{\alpha} \\
& \geq \frac{\frac{k\cutoff }{2}(2-\alpha)  +\ell \cutoff  }{\alpha}
\geq \frac{\frac{\ell \cutoff}{2(\alpha-1)}(2-\alpha) +\ell \cutoff  }{\alpha}
= \frac{\ell \cutoff(2-\alpha) +2\ell \cutoff(\alpha-1)  }{2\alpha(\alpha-1)}
 = \frac{\ell \cutoff	}{2(\alpha-1)}
\geq \frac{k\cutoff	}{2},
\end{align*}
as needed.
\end{enumerate}
This completes the argument.
\EPF

Since we want to set the same price for all items, we need a way to compare the welfare obtained from setting different prices in order to apply Lemma~\ref{l_kv2}. The next lemma states that the number of items that an agent buys can only decrease as the price goes up.

\BL \label{l_dec_price}
Let $\val$ be any non-decreasing valuation function, and let $p_1 < p_2$ be two prices (assigned to all items). Let $x_1 = \argmax (\val(x)-p_1\cdot x)$, and $x_2 = \argmax (\val(x)-p_2\cdot x)$.
Then $x_1 \geq x_2$.
\EL

\BPF
We know that $\val(x_1) - x_1 \cdot p_1 \geq \val(x_2) - x_2 \cdot p_1$ and $\val(x_2) - x_2 \cdot p_2 \geq \val(x_1) - x_1 \cdot p_2$. Using these inequalities gives us:
\begin{eqnarray*}
 & & \frac{\val(x_1)-\val(x_2)}{p_1} \geq x_1-x_2 \geq \frac{\val(x_1)-\val(x_2)}{p_2}\\
 & \implies &  (p_2-p_1)(\val(x_1)-\val(x_2))  \geq 0 \\
 & \implies &  \val(x_1)-\val(x_2)  \geq 0 \\
  & \implies &  \val(x_1) \geq \val(x_2).
\end{eqnarray*}
If it were the case that $x_1<x_2$, then since $\val$ is non-decreasing, we must have $\val(x_1)=\val(x_2)$. Our initial inequalities then imply that $x_1=x_2$, a contradiction. So $x_1\geq x_2$, as claimed.
\EPF

By combining Lemmas~\ref{l_kv2} and \ref{l_dec_price}, we can derive a guarantee on the social welfare obtained by setting a uniform price.

\BL \label{l_mv2}
Let $\val_1,\ldots, \val_n$ be the subadditive valuations of the agents, and let $\valtwo_1, \ldots ,\valtwo_n $ be the respective minimal submodular functions such that $\valtwo_i \geq \val_i$, as defined in Proposition~\ref{l_submod}. Define $\cutoff$ as in Section \ref{sec:prelim} with respect to the $\valtwo_i$'s.
When offering the agents the price $\cutoff/2$, the social welfare obtained will be at least $m\cutoff/2$.
\EL

\BPF
If all items were sold, then the revenue will be $m\cutoff/2$ and therefore the social welfare will be at least $m\cutoff/2$ as needed.
Otherwise, let $\ell_i= \max \{x\mid(\valtwo_i(x)-\valtwo_i(x-1)) \geq \cutoff\}$. We know that there exists an agent $i$ such that  she bought less than $\ell_i$ items (else all items would have been sold).
Suppose that upon the arrival of agent $i$ there are $k$ items remaining for sale.
If $k \geq \ell_i $ then agent $i$ will buy at least $\ell_i$ items, contradicting our assumption, so $k<\ell_i$.

Let $\cutoff'$ be the $k$th marginal value of $\valtwo_i$; in particular, $\cutoff'\geq\cutoff$. By Lemma \ref{l_kv2}, if we set a price of $\cutoff'/2$, then the welfare obtained from agent $i$ is at least $k\cutoff'/2$. Now, by Lemma \ref{l_dec_price}, we can only sell more items to agent $i$ by setting a lower price $\cutoff/2$. This implies that the welfare obtained by this agent is at least $k\cutoff/2$.
In addition, it is known that $m-k$ items were sold prior to the arrival of agent $i$, contributing at least $(m-k)\cutoff/2$ to the social welfare.
We conclude that the total social welfare is at least $(m-k)\cutoff/2 + k\cutoff/2 = m\cutoff/2$.
\EPF

With Lemma~\ref{l_mv2} in hand, we can now derive the $1/3$ approximation on the social welfare with uniform pricing.

Let $\val_1,\ldots, \val_n$ be the subadditive valuations of the agents, and let $\valtwo_1, \ldots ,\valtwo_n $ be the respective minimal submodular functions such that $\valtwo_i \geq \val_i$, as defined in Proposition~\ref{l_submod}. Define $V,\cutoff,m'$, and $\epsilon$ with respect to the $\valtwo_i$'s as in Section \ref{sec:prelim}. We show that one of the following pricing schemes guarantees at least $1/3$ of the optimal social welfare:

(P1) a uniform price of $\cutoff+ \epsilon$;

(P2) a uniform price of $\cutoff/2$.

Let $(x_1,\ldots ,x_n)$ be the optimal partition of items to agents (where $x_i$ is the number of items allocated to agent $i$), and let $OPT= \val_1(x_1) +\ldots + \val_n(x_n)$ be the optimal social welfare.
It is easy to see that $OPT \leq \sum_{\cutoff<\cutoff' \in V}\cutoff'  + (m-m')\cutoff$ (since the RHS is the optimal allocation under valuations $\valtwo_i$'s, which is at least as high as the optimal welfare of $\val_i$'s).

Pricing (P1) guarantees exactly a welfare of $\sum_{\cutoff<\cutoff' \in V}\cutoff'$.
Pricing (P2) by Lemma \ref{l_mv2} guarantees a welfare of at least $m\cutoff/2$.
The best of the two guarantees yields at least $1/3$ of the optimal social welfare.
\EPF

If there are two identical agents, the bound in Theorem~\ref{t_subadd_3} can be improved to $2/3$.

\BT \label{t_subadd_2_iden}
For every market of $m$ items and two identical agents with symmetric subadditive valuations, there exists a uniform static item pricing $\prices$ that guarantees at least $2/3$ of the optimal social welfare.
\ET

\BPF
Let $OPT$ be the optimal social welfare, and $\val$ the valuation function of the two agents. Set a uniform price of $OPT/3m$.  In the optimal allocation, the total utility of the two agents, if they were to buy the items at this price, is $OPT - m\cdot OPT/3m = 2OPT/3$, so at least one agent gets utility at least $OPT/3$. This means that in the allocation that results from setting the price $OPT/3m$, the first agent must also get utility at least $OPT/3$, since all items are still available. If the first agent buys at most half of the items, so does the second agent, and the total utility (and hence the welfare) is at least $2OPT/3$ as we want. So suppose that the first agent buys $m' > m/2$ items. The first agent contributes $m'\cdot OPT/3m$ to the revenue and has utility at least $OPT/3$, so since welfare is the sum of revenue and utility, we have
$\val(m') \geq \frac{OPT}{3}\cdot\left(1 + \frac{m'}{m}\right)$.

It suffices to show that the second agent contributes at least $2OPT/3 - \val(m')$ to the welfare. We consider two cases.

\begin{enumerate}
\item $\val(m-m') \geq \frac{2OPT}{3} - \val(m') + \frac{OPT}{3m}\cdot(m-m')$. In this case, the second agent can guarantee utility at least $2OPT/3 - \val(m')$ by buying all of the remaining $m-m'$ items, so the agent's actual utility (and hence contribution to the welfare) is at least $2OPT/3 - \val(m')$.

\item $\val(m-m') < \frac{2OPT}{3} - \val(m') + \frac{OPT}{3m}\cdot(m-m')$. Let $k:=\left\lfloor\frac{m'}{m-m'}\right\rfloor$, and let $x := m'-k(m-m')=(k+1)m' - km\geq 0$. By considering the possibility that the second agent buys $x$ items, we get the desired conclusion if $\val(x) \geq \frac{2OPT}{3} - \val(m') + \frac{OPT}{3m}\cdot x$. Hence we may assume that $\val(x) < \frac{2OPT}{3} - \val(m') + \frac{OPT}{3m}\cdot x$. We have
\begin{align*}
\val(m')
&= \val(k(m-m') + x) \\
&\leq k\cdot \val(m-m') + \val(x) \\
&< (k+1)\left(\frac{2OPT}{3} - \val(m')\right) + \frac{OPT}{3m}\cdot m',
\end{align*}
where the first inequality follows from subadditivity of $v$.
Rearranging, we find that
$(k+2)\cdot \val(m') < \frac{OPT}{3}\cdot \left(2k+2+\frac{m'}{m}\right)$.
Combining this with the previous estimate $\val(m') \geq \frac{OPT}{3}\cdot\left(1 + \frac{m'}{m}\right)$, we have
$(k+2)\left(1+\frac{m'}{m}\right) < 2k+2+\frac{m'}{m}$,
which rearranges to $x<0$, a contradiction. Hence the second case cannot occur.
\end{enumerate}
This completes the proof.
\EPF

\subsection{Lower bounds}

The next propositions show that the bound in Theorem~\ref{t_subadd_3} cannot be improved to more than $1/2$, and in the case of using only uniform pricing, cannot be improved to more than $1/3$. This means that the bound in Theorem~\ref{t_subadd_3} is tight for uniform pricing.

\BP
\label{p_subadd_1_2}
There is a market with symmetric subadditive valuations with $m$ items and two agents such that no static pricing $\prices$ guarantees more than $1/2$ of the optimal social welfare.
\EP

\BPF
Consider a market where the first agent has a valuation $\vali[1](i) = 1$ for $ i \leq m-1$ and $\vali[1](m)=2$, and the second agent has a unit-demand valuation with value $\frac{1}{m-1}$.
The optimal allocation is giving all the items to the first agent (therefore the optimal social welfare is $2$).
Let  $\prices$ be any pricing.
If there exists $i$ such that $p_i \leq \frac{1}{m-1}$, then if agent 2 arrives first, she will buy an item and the social welfare will be bounded by $1+\frac{1}{m-1}$.
Otherwise we have $p_i > \frac{1}{m-1}$ for all $i$. In this case, agent 1 has no incentive to buy more than one item, and again the social welfare will be bounded by $1+\frac{1}{m-1}$. Taking $m$ large, this yields the ratio of $1/2$ as claimed.
\EPF

\BP
\label{p_subadd_1_3}
There is a market with symmetric subadditive valuations with $m$ items and three agents such that no uniform static pricing $\prices$ guarantees more than $1/3$ of the optimal social welfare.
\EP

\BPF
Consider a market where the first agent has a valuation $\vali[1](i) = m-1+i$ for $ i \leq m-1$ and $\vali[1](m)=3m-2$, the second agent has a unit-demand valuation with value $2$, and the third agent has an additive valuation $\vali[3](i) = i$.
The optimal allocation is giving all items to the first agent, which yields the optimal social welfare of $3m-2$.
Let  $p$ be any uniform pricing.
If  $p \leq 1$, then if agent 3 arrives first, she will buy all the items and the social welfare will be bounded by $m$.
If  $1< p \leq 2$, then if agent 2 arrives first, then she will buy an item and then only the first agent will buy an additional item, and the social welfare will be bounded by $m+2$.

Otherwise we have that $p >2$. In this case, agent 1 has no incentive to buy more than one item, the other agents will not buy any item, and again the social welfare will be bounded by $m$. Taking $m$ large, this yields a ratio of $1/3$, as claimed.
\EPF

In the case of two identical agents, the approximation cannot be improved to more than $3/4$. In this special case, we can guarantee at least half of the social welfare by applying Theorem~\ref{t_general_2}.

\BP \label{t_3_4_subadditive_identical}
There exists a market of $m$ items and two identical agents with a subadditive valuation such that no static pricing guarantees more the $3/4$ of the optimal social welfare.
\EP

\BPF
Assume that $m$ is even for simplicity.
Consider the function $$
\val(i)=
\begin{cases}
m/4 & i < m/2\\
m/2 -1-\epsilon & i =m/2\\
m/2 & i > m/2\\
\end{cases}.
$$
And consider a market where there are two agents with valuation $\val$.
The optimal allocation is to give $m/2$ to each agent, which generates a social welfare of $m-2-2\epsilon$. In order to set prices so that each agent will want to buy exactly $m/2$ items, we need that the $m/2$ more expensive items cost less than $m/2 -1-\epsilon$, and therefore there exists one among them that cost less than $1$.
When the first agent arrives, she will buy either exactly $m/2+1$ items or less than $m/2$ items, since the cheapest $m/2+1$ cost less than $1$ each. Therefore the social welfare obtained by any static pricing in this market cannot be more than $3m/4 \approx 3OPT/4$, as claimed.
\EPF

If we use uniform pricing, we cannot guarantee more than $2/3$ of the welfare for two identical agents. This means that the bound in Theorem~\ref{t_subadd_2_iden} is tight.

\BP
\label{p_subadd_2_3_iden}
There is a market with symmetric subadditive valuations with $m$ items and two identical agents such that no uniform static pricing $\prices$ guarantees more than $2/3$ of the optimal social welfare.
\EP

\BPF
Consider a market where both agents have the valuation $v(i)=m+i$. An optimal allocation is to give one item to the first agent and the remaining items to the second agent for a social welfare of $3m$. Let $p$ be any uniform pricing. If $p\leq 1$, then the first agent buys all the items and the social welfare is $2m$. Otherwise we have $p>1$, and both agents only buy one item for a social welfare of $2m+2$. Taking $m$ large, this yields a ratio of $2/3$ as claimed.
\EPF

\section{General valuations}

In this section we consider general valuations.
While the analysis assumes monotonicity, all results hold even for non-monotone valuations; simply do all calculations based on the monotone closure of the valuations, and the results hold for the valuations themselves.

\subsection{Worst-case ordering}

We first show that for general valuations, we cannot guarantee more than $1/m$ of the optimal welfare even if we know the order of arrival.
	
\BP
\label{p_general_lower}
There is a market with symmetric valuations over $m$ items and two agents such that no static pricing $\prices$ guarantees more than $1/m$ of the optimal social welfare even for a known order of arrival.
\EP

\BPF
Consider a market where the first agent has a unit-demand valuation with value $1$, and the second agent is single-minded for the set of all items with value $m$.
The optimal allocation is giving all the items to the second agent (therefore the optimal social welfare is $m$).
Let  $\prices$ be any pricing.
If there exists $i$ such that $p_i < 1$, then the first agent will buy an item and the social welfare will be bounded by $1$.
Otherwise, $p_i \geq 1$ for all $i$ and therefore the price of the bundle of all items is at least $m$. This means agent 2 has no incentive to buy the bundle and the social welfare obtained by these prices is $0$.
\EPF

The example in Proposition~\ref{p_general_lower} is as bad as it gets: We can guarantee $1/m$ of the optimal welfare with uniform pricing.

\BT
\label{t_general_upper}
For every market of $m$ identical items, there exists a uniform static item pricing $\prices$ that guarantees at least $1/m$ of the optimal social welfare.
\ET

\BPF
Consider an allocation that yields the optimal welfare, and let $\beta$ be the highest average welfare per item among all agents in this allocation. Clearly, $OPT\leq m\beta$. Set the uniform price to be $\beta-\epsilon$ for small enough $\epsilon$. Then at least one item will be sold, since the agent with the highest average welfare per item in the optimal allocation can achieve positive utility. It follows that the welfare obtained by this pricing is at least $\beta$.
\EPF

\subsection{Best-case ordering}

Next, we show that if we can choose the order of arrival, then we can guarantee at least half of the optimal welfare. We remark that when agents are identical, the order of arrival does not matter, and therefore our result holds for the setting with identical agents as well.

\BT \label{t_general_2}
For every market of $m$ identical items, there exists a uniform static item pricing $\prices$ along with an order of arrival that guarantees at least $1/2$ of the optimal social welfare.
\ET

\BPF
Let $\val_1,\ldots, \val_n$ be the valuations of the agents, and let $\valtwo_1, \ldots ,\valtwo_n $ be the respective minimal submodular functions such that $\valtwo_i \geq \val_i$, as defined in Proposition~\ref{l_submod}. Define $V,\cutoff,$ and $m'$ as in Section \ref{sec:prelim} with respect to the $\valtwo_i$'s.
In addition, one can find a sufficiently small $\epsilon$ so that at the price $\cutoff -\epsilon$, whenever an agent buys an additional set of items that belong to the same ``marginal value segment'' of $\valtwo_i$, the average marginal value of items in the set is at least $\cutoff$. Let $(x_1,\ldots ,x_n)$ be the optimal allocation of items to agents where $x_i$ is the number of items allocated to agent $i$, and let $OPT= \val_1(x_1) +\ldots + \val_n(x_n)$ be the optimal social welfare.

We show how to construct a pricing scheme that guarantees at least $OPT/2$:

(P1) a uniform price of $\cutoff+ \epsilon$ with any arrival order (in this case the order does not influence the social welfare);

(P2) a uniform price of $\cutoff- \epsilon$ with some arrival order $\sigma$ to be chosen.\\
It is easy to see that $OPT \leq \sum_{\cutoff<\cutoff' \in V}\cutoff'  + (m-m')\cutoff$, since the RHS is the optimal allocation under valuations $\valtwo_i$'s, which is at least as high as the optimal welfare under $\val_i$'s.

Pricing (P1) guarantees a welfare of exactly $\sum_{\cutoff<\cutoff' \in V}\cutoff'$.
We next show that we can generate orders $\sigma,\sigma'$ so that the social welfare of the best of pricing (P1) and pricing (P2) with order $\sigma$ or $\sigma'$ will guarantee at least $OPT/2$. We start with an arbitrary order $\sigma$ for (P2) and consider three cases.

\begin{enumerate}
	\item All items are sold. The social welfare will be at least $m\cutoff$, and the best of pricing (P1) and (P2) with order  $\sigma$ will generate a social welfare of at least $OPT/2$.
	
	\item Not all items are sold, but the number of items bought by every agent $i$ is at least the number of items with marginal contribution greater than $\cutoff$ (with respect to $\valtwo_i$). Suppose that $m' \leq k<m$ items are sold in total (and therefore the social welfare of pricing (P2) and order $\sigma$ is exactly  $\sum_{\cutoff<\cutoff' \in V}\cutoff' + (k-m')\cutoff$).
	In this case we have that there exists an agent $i$ such that $\valtwo_i(m-k+1)- \valtwo_i(m-k) \geq \cutoff$; otherwise the agents would get all the items that contribute to them at least $\cutoff$ until at least $k+1$ items are sold.
	We set $\sigma'$ to be an order where agent $i$ is the first agent to arrive.
	In order  $\sigma'$ and pricing (P2), we have that the first agent (i.e., agent $i$) will buy at least $m-k+1$ items (since the marginal value of the first $m-k+1$ items is at least $\cutoff$ and the price is less than $\cutoff$, and all items are available in this stage).
	The sum of the social welfare of the two schemes (with order $\sigma$ or $\sigma'$) is at least $\sum_{\cutoff<\cutoff' \in V}\cutoff' + (k-m')\cutoff + (m-k+1)\cutoff \geq OPT$, and therefore the best of the two will generate at least $OPT/2$ as needed.
	
	\item Not all items are sold, and the number of items bought by some agent $i$ is less than the number of items with marginal contribution greater than $\cutoff$ (with respect to $\valtwo_i$). Suppose that $k$ items are sold in total.
	In this case we have that $k> m-m'$; otherwise, when an agent with marginal value greater than $\cutoff$ is offered, there will be enough items available for her to buy all items with marginal value greater than $\cutoff$ (since there are at most $m'$ such items and at least $m'$ items are available for her to buy).
	Therefore the social welfare of this pricing scheme is at least $k\cutoff > (m-m')\cutoff$.
	Therefore the best of this pricing scheme and pricing scheme (P1) will generate a social welfare of at least $OPT/2$.	
\end{enumerate}
This concludes the argument.
\EPF

The bound $1/2$ in Theorem~\ref{t_general_2} is also tight.

\BP \label{p_general_2_upper}
There exists a market of $m$ items for which no static pricing and order of arrival yields more than $1/2$ of the optimal social welfare.
\EP

\BPF
Assume that $m$ is even for simplicity. Consider the function $$
\val(i)=
\begin{cases}
0 & i < m/2\\
m/2 -\epsilon & i =m/2\\
m/2 +1& i > m/2\\
\end{cases}.
$$
And consider a market where there are two agents with valuation $\val$.
Since the agents have the same valuation, the order of arrival does not influence the social welfare.
The optimal allocation is to give $m/2$ items to each agent. In order to set prices so that each agent will want to buy at least $m/2$ items, we need that the $m/2$ more expensive items together cost less than $m/2 -\epsilon$, and therefore there exists one among them that costs less than $1$.
When the first agent arrives, she will buy exactly $m/2+1$ items, since the cheapest $m/2+1$ cost less than $1$ each. Therefore the social welfare obtained by any static pricing in this market cannot exceed $m/2+1 \approx OPT/2$, as claimed.
\EPF

\section{Bayesian setting}

In this section, we consider the Bayesian setting, where the valuation function of each agent is drawn independently from a distribution which can be different for different agents.

\subsection{XOS valuations}

Feldman et al. \cite{FGL15} showed that if agents' valuations are drawn independently from a distribution over XOS valuation functions, then there exist prices that yield expected welfare at least half of the expected optimal welfare.
These posted prices are non-uniform, even when the result is specialized to the case of identical items.
We first restate Feldman et al.'s result and then show that if the items are identical, then the same bound can be obtained using uniform prices.

\BT \label{t_xos_2}\cite{FGL15}
Let $\F=F_1\times \ldots \times F_n$ be a product distribution over XOS valuation functions. For every $\vals =(\val_1,\ldots,\val_n)\in \F$, let $X^*(\vals)=(X^*_1(\vals),\ldots,X^*_n(\vals))$ be any allocation that maximizes the social welfare.
Let $\textbf{a}=(a_1,\ldots,a_n)$ be additive functions such that $\val_i(S)\geq a_i(S)$ for any subset $S$ of items, and $\val_i(X^*_i)=a_i(X^*_i)$.
When the items are offered at prices $p_j= E_{\vals\in \F}[a_i(j)/2 \mbox{ where } j \in X^*_i(\vals)]$,
the expected social welfare is at least $OPT/2$.
\ET

\BT
\label{t_xos_uniform}
Let $\F=F_1 \times \ldots \times F_n$ be a product distribution over symmetric XOS valuation functions. Let $OPT$ be the expected optimal social welfare.
When all items are offered at the uniform price $OPT/2m$, the expected social welfare is at least $OPT/2$.
\ET

\BPF
Let $G$ be the uniform distribution over $m$ different classes of the zero valuation. We notate the $m$ different classes as $0_\sigma$ for $\sigma \in [m]$.
Let $\hat{\F}$ be the distribution $\F \times G$. In other words, in addition to the distribution $\F$ over the valuations of the $n$ agents, we also add another agent whose valuation is always zero. Clearly, the distributions $\hat{\F}$ and $\F$ have the same expected optimal social welfare, and for any prices $\prices>0$, the expected social welfare when considering $\hat{\F}$ and $\F$ is the same since the added agent never buys an item.

Let $\vals=(\val_1,\ldots,\val_n) \in \F$, and let $X^*(\vals)=(X^*_1(\vals),\ldots,X^*_n(\vals))$ be any optimal allocation with respect to $\vals$.
We define an optimal allocation for every
$\hat{\vals} = (\val_1,\val_2,\ldots , \val_n, 0_\sigma) \in \hat{\F}$ to be
$\hat{X}_i^*(\hat{\vals}) = \{j\in[m] \mid j+\sigma \in X_i^*(\vals) \}$, where $j+\sigma$ is considered modulo $m$.
This is an optimal allocation since the functions are symmetric and the allocation gives the same number of items to each agent as the original optimal allocation.
Let $\textbf{a}=(a_1,\ldots,a_n)$ be the additive functions corresponding to $X^*(\vals)$ that maximize the social welfare, as defined in Theorem~\ref{t_xos_2}.
Let $\hat{\textbf{a}}= (\hat{a}_1,\ldots,\hat{a}_n)$ be the additive functions where $\hat{a}_i(\{j\}) = a_{i}(\{j+\sigma\})$. Due to the symmetry over the items, these functions also maximize the social welfare.

We now show that with the new distribution $\hat{\F}$, the prices are all identical and equal to $OPT/2m$.
The prices calculated according to Theorem \ref{t_xos_2} are
\begin{eqnarray*}
	p_j & = & E_{\hat{\vals} \sim \hat{\F}}\left[\hat{a}_i(\{j\})/2 \mbox{ where } j \in \hat{X}^*_i(\hat{\vals})\right]
	=  E_{\vals \sim \F , 0_\sigma \sim G}\left[a_i(\{j+\sigma\})/2 \mbox{ where } j + \sigma \in X^*_i(\vals)\right] \\
	& = & \frac{1}{2m} \cdot E_{\vals \sim \F}\left[\sum_{j'} a_i(\{j'\}) \mbox{ where } j' \in X^*_i(\vals)\right]
	=  \frac{1}{2m} \cdot OPT(\vals) = \frac{OPT}{2m}.
\end{eqnarray*}
This concludes the argument.
\EPF

\subsection{Subadditive and general valuations}

We now define a notion that describes how close an arbitrary valuation function is to an XOS function and derive approximation results in terms of this closeness quantity. The proof of Theorem \ref{t_c_xos} follows the analysis presented by Feldman et al. \cite{FGL15}.

\BD
We say that a (not necessarily symmetric) valuation function $\val$ is \emph{$c$-close} to XOS if there exists an XOS function $\valtwo$  such that for every set of item $S$, it holds that $\val(S)/c \leq \valtwo(S) \leq \val(S)$.
\ED

\BT \label{t_c_xos}
For any product distribution $\F$ over (not necessarily symmetric) valuation functions  that are $c$-close to XOS, there exist anonymous prices  $\prices$ that guarantee an expected social welfare of at least $1/2c$ of the optimal expected welfare.
\ET

\BPF
For every $\vals=(\val_1,\ldots,\val_n)$, let $X^*(\vals)=(X^*_1,\ldots,X^*_n)$ be an allocation that maximizes the social welfare with respect to the functions $\vals$.
Let $\valtwos = (\valtwo_1, \ldots, \valtwo_n)$ be the vector of the corresponding XOS functions; i.e.,  $\val_i/c\leq \valtwo_i\leq \val_i$ for every $i$.
Let $\textbf{a}=(a_1,\ldots,a_n)$ be additive functions such that $\valtwo_i(S)\geq a_i(S)$ for any subset $S$ of items, and $\valtwo_i(X_i^*)=a_i(X_i^*)$.

For every item $j$, we define $w_j$	to be the social welfare contribution of item $j$ under valuations $\textbf{a}$, and we set its price to be $p_j=E_{\vals\sim \F}[w_j]/2$. (Recall that every vector of valuations $\vals$ defines an optimal allocation $X^*(\vals)$ and XOS functions $\valtwos$, and the allocation and the XOS functions in turn define additive functions $\textbf{a}$). The expected maximal social welfare is $E[\SW] = E_{\vals\sim \F} \left[\sum \val_i(X^*_i)\right]$.

Fix $i$ and $(\val_i, \vals_{-i})$. Let $SOLD_i (\vals)$ denote the items that have been sold before the arrival of agent $i$.
Consider $\vals'_{-i} \sim \F_{-i}$ which is independent of $\vals$, and let $\valtwos'_{-i}$ be the XOS functions that approximate  $\vals'_{-i}$.
Let $X^*(\val_i,\vals'_{-i})$ be an optimal allocation under the valuations $(\val_i,\vals'_{-i})$; in particular, $X^*_i(\val_i,\vals'_{-i})$ is the bundle assigned to agent $i$ in this allocation.
Define $\textbf{a}'$ analogously to $\textbf{a}$ with respect to the allocation $X^*(\val_i,\vals'_{-i})$ and the valuations $\valtwos'_{-i}$.
Let $S_i(\val_i,\vals_{-i},\vals'_{-i})= X^*_i(\val_i,\vals'_{-i}) \backslash SOLD_i(\vals)$ be the subset of items in  $X^*_i(\val_i,\vals'_{-i})$ that are available when agent $i$ arrives; therefore her utility is at least the utility from purchasing this set of items.
Thus her utility satisfies
$$u_i(\vals)  \geq E_{\vals'_{-i}}\left[\sum_{j \in S_i(\val_i,\vals_{-i},\vals'_{-i})} (a'(\{j\})-p_j)^+\right],$$
where $f^+=\max(f,0)$.
Here we use the assumption that $a_i'(S)\leq \valtwo_i(S)\leq \val_i(S)$ for any subset $S$ of items.
Adding these inequalities for all $i$ and taking expectation over $\vals$ gives us:

\begin{eqnarray*}
 E_{\vals\sim \F}\left[\sum_{i=1}^n u_i(\vals)\right] & \geq &
\sum_{j\in M} \sum_{i=1}^n E_{\val_i,\vals_{-i},\vals'_{-i}}[1_{j \in X^*_i(\val_i,\vals'_{-i})} \cdot (a'(\{j\})-p_j)^+ \cdot 1_{j \not \in SOLD_i(\vals)}] \\
&= & \sum_{j\in M} \sum_{i=1}^n E_{\val_i,\vals_{-i},\vals'_{-i}}[1_{j \in X^*_i(\val_i,\vals'_{-i})} \cdot (a'(\{j\})-p_j)^+ \cdot 1_{j\not \in SOLD_i(\vals_{-i})}] \\
&= & \sum_{j\in M} \sum_{i=1}^n E_{\vals_{-i}}[1_{j\not \in SOLD_i(\vals_{-i})}]\cdot E_{\val_i,\vals'_{-i}}[1_{j \in X^*_i(\val_i,\vals'_{-i})} \cdot (a'(\{j\})-p_j)^+] \\
&\geq & \sum_{j\in M} \sum_{i=1}^n Pr_{\vals}[j\not \in SOLD(\vals)] \cdot  E_{\val_i,\vals'_{-i}}[1_{j \in X^*_i(\val_i,\vals'_{-i})} \cdot (a'(\{j\})-p_j)^+ ] \\
&= & \sum_{j\in M} \left(Pr_{\vals}[j\not \in SOLD(\vals)] \cdot  \sum_{i=1}^n E_{\vals}[1_{j \in X^*_i(\vals)} \cdot (a(\{j\})-p_j)^+ ] \right)\\
&\geq & \sum_{j\in M} \left(Pr_{\vals}[j\not \in SOLD(\vals)] \cdot  \sum_{i=1}^n E_{\vals}[1_{j \in X^*_i(\vals)} \cdot (a(\{j\})-p_j) ]\right) \\
&= & \sum_{j\in M} Pr_{\vals}[j\not \in SOLD(\vals)] \cdot p_j. \\
\end{eqnarray*}

In the second line, we replace $SOLD_i(\vals)$ by $SOLD_i(\vals_{-i})$ since the set of items sold before the arrival of agent $i$ does not depend on $\val_i$. In the fourth line, we decrease each probability $Pr_{\vals_{-i}}[j\not \in SOLD(\vals_{-i})]$ to $Pr_{\vals}[j\not \in SOLD(\vals)]$; the inequality holds since all the remaining terms are non-negative. In the fifth line, we replace $\vals'_{-i}$ by $\vals_{-i}$ since these valuations are drawn from the same distribution and we are taking their expectation. The last line follows from the definition of $p_j$.

The expected revenue of this mechanism is
$\sum_j Pr_{\vals}[j \in SOLD(\vals)] \cdot p_j $.
Therefore the expected social welfare guaranteed by these prices is at least
$\sum_{j\in M} Pr_{\vals}[j \in SOLD(\vals)] \cdot p_j + \sum_{j\in M} Pr_{\vals}[j\not \in SOLD(\vals)] \cdot p_j \\
= \sum_{j\in M} p_j.$
Finally, $$\sum_{j\in M} p_j = \frac{1}{2}\cdot E_{\vals}\left[\sum_{j\in M} w_j\right]=\frac{1}{2}\cdot E_{\vals}\left[\sum_{i=1}^n \valtwo_i(X^*_i)\right] \geq \frac{1}{2c}\cdot E_{\vals}\left[\sum \val_i(X^*_i)\right],$$ and therefore we have a $2c$ approximation, as claimed.
\EPF

Since any symmetric subadditive function is 2-close to XOS according to Proposition \ref{subadd-xos-bound}, Theorem \ref{t_c_xos} implies that we can obtain at least $1/4$ of the expected optimal welfare when the agents' valuations are drawn from a product distribution over subadditive valuations. In addition, using techniques similar to those in the proof of Theorem \ref{t_xos_uniform}, we can achieve this with uniform prices.

\BT
\label{t_subadd_uniform}
Let $\F=F_1 \times \dots \times F_n$ be a product distribution over symmetric subadditive valuation functions. Let $OPT$ be the expected maximal social welfare.
There exists a uniform price on the items for which the expected social welfare is at least $OPT/4$.
\ET

Finally, we show that we cannot extend our results in the Bayesian setting to general valuations, even if we can control the order of arrival.

\BP
\label{p_bayesian_lower}
There is a market with a distribution over symmetric valuations with $n$ agents and $m=n^2$ items such that no static pricing $\prices$ yields expected welfare more than $\Theta(1/n)$ of the optimal expected welfare, even if we can control the arrival order.
\EP

\BPF
Consider a market with $n$ identical agents (so that the arrival order does not affect the expected welfare). Each agent is unit-demand with value 1 with probability $1-1/n$, and single-minded with value $m$ over the grand bundle with probability $1/n$. With probability at least $1-1/e$, there exists a single-minded agent, implying that the expected optimal welfare is at least $(1-1/e)m=\Theta(n^2)$. If we price all items at least 1, then a single-minded agent will not buy anything, and the welfare is at most $n$. Otherwise, we price some item below 1. With probability $1-1/n$, the first agent will buy this item and the welfare is again at most $n$. So the expected welfare in this case is at most $(1-1/n)n+(1/n)m=O(n)$.
\EPF

\section{Discussion}

In this paper, we study the fraction of the optimal social welfare that can be achieved via posted prices in markets with identical items under various assumptions on the designer's information and agents' valuations. We show that in the Bayesian setting, uniform posted prices can guarantee $1/2$ and $1/4$ of the optimal welfare for XOS and subadditive valuations, respectively. If the designer has full information on agents' valuations, then $1/3$ of the optimal welfare can be obtained via uniform prices for subadditive valuations. For general valuations, we exhibit a tight bound of $1/m$ for both uniform and non-uniform prices; on the other hand, if the designer can control the arrival order, then $1/2$ of the optimal welfare can be guaranteed for such valuations.

Our work sheds light on the power of uniform prices for settings with identical items. For submodular valuations in the full-information setting, there is a gap between the guarantee that can be obtained by uniform and non-uniform prices, while for XOS valuations in the Bayesian setting there is no gap. It remains open whether such a gap exists for subadditive valuations, both for the full-information and the Bayesian setting. Another important open question in the more general setting with non-identical (i.e., heterogeneous) items that remains from previous work is whether the bound of $\log m$ for subadditive valuations by Feldman et al. \cite{FGL15} can be improved, or whether a constant approximation can be achieved in this setting.

\section*{Acknowledgments}
The work of M. Feldman and T. Ezra was partially supported by the European Research Council under the European Union's Seventh Framework Programme (FP7/2007-2013) / ERC grant agreement number 337122, and by the Israel Science Foundation (grant number 317/17).

\bibliographystyle{plain}
\bibliography{pricing-multi-unit}

\begin{thebibliography}{10}

\bibitem{Ausubel04}
Lawrence~M. Ausubel.
\newblock An efficient ascending-bid auction for multiple objects.
\newblock {\em American Economic Review}, 94(5):1452--1475, 2004.

\bibitem{BBDS11}
Moshe Babaioff, Liad Blumrosen, Shaddin Dughmi, and Yaron Singer.
\newblock Posting prices with unknown distributions.
\newblock In {\em Proceedings of the 1st Innovations in Computer Science},
  pages 166--178, 2011.

\bibitem{BILW14}
Moshe Babaioff, Nicole Immorlica, Brendan Lucier, and S.~Matthew Weinberg.
\newblock A simple and approximately optimal mechanism for an additive buyer.
\newblock In {\em Proceedings of the 55th {IEEE} Annual Symposium on
  Foundations of Computer Science}, pages 21--30, 2014.

\bibitem{BDHS15}
MohammadHossein Bateni, Sina Dehghani, MohammadTaghi Hajiaghayi, and Saeed
  Seddighin.
\newblock Revenue maximization for selling multiple correlated items.
\newblock In {\em Proceedings of the 23th Annual European Symposium on
  Algorithms}, pages 95--105, 2015.

\bibitem{BhawalkarR12}
Kshipra Bhawalkar and Tim Roughgarden.
\newblock Simultaneous single-item auctions.
\newblock In {\em Proceedings of the 8th International Workshop on Web and
  Internet Economics}, pages 337--349, 2012.

\bibitem{BlumrosenH08}
Liad Blumrosen and Thomas Holenstein.
\newblock Posted prices vs. negotiations: an asymptotic analysis.
\newblock In {\em Proceedings of the 9th {ACM} Conference on Electronic
  Commerce}, page~49, 2008.

\bibitem{CHK07}
Shuchi Chawla, Jason~D. Hartline, and Robert~D. Kleinberg.
\newblock Algorithmic pricing via virtual valuations.
\newblock In {\em Proceedings of the 8th {ACM} Conference on Electronic
  Commerce}, pages 243--251, 2007.

\bibitem{CHMS10}
Shuchi Chawla, Jason~D. Hartline, David~L. Malec, and Balasubramanian Sivan.
\newblock Multi-parameter mechanism design and sequential posted pricing.
\newblock In {\em Proceedings of the 42nd {ACM} Symposium on Theory of
  Computing}, pages 311--320, 2010.

\bibitem{CMS10}
Shuchi Chawla, David~L. Malec, and Balasubramanian Sivan.
\newblock The power of randomness in {B}ayesian optimal mechanism design.
\newblock In {\em Proceedings of the 11th {ACM} Conference on Electronic
  Commerce}, pages 149--158, 2010.

\bibitem{CKS08}
George Christodoulou, Annam{\'{a}}ria Kov{\'{a}}cs, and Michael Schapira.
\newblock Bayesian combinatorial auctions.
\newblock In {\em Proceedings of the 35th International Colloquium on Automata,
  Languages and Programming}, pages 820--832, 2008.

\bibitem{CEFJ15}
Ilan~Reuven Cohen, Alon Eden, Amos Fiat, and Lukasz Jez.
\newblock Pricing online decisions: Beyond auctions.
\newblock In {\em Proceedings of the 26th Annual {ACM-SIAM} Symposium on
  Discrete Algorithms}, pages 73--91, 2015.

\bibitem{CEFF16}
Vincent Cohen{-}Addad, Alon Eden, Michal Feldman, and Amos Fiat.
\newblock The invisible hand of dynamic market pricing.
\newblock In {\em Proceedings of the 2016 {ACM} Conference on Economics and
  Computation}, pages 383--400, 2016.

\bibitem{DFKL17}
Paul D{\"{u}}tting, Michal Feldman, Thomas Kesselheim, and Brendan Lucier.
\newblock Posted prices, smoothness, and combinatorial prophet inequalities.
\newblock In {\em Proceedings of the 58th IEEE Annual Symposium on Foundations
  of Computer Science}, pages 540--551, 2017.

\bibitem{EOS07}
Benjamin Edelman, Michael Ostrovsky, and Michael Schwarz.
\newblock Internet advertising and the generalized second-price auction:
  Selling billions of dollars worth of keywords.
\newblock {\em The American Economic Review}, 97(1):242--259, 2007.

\bibitem{EFFTW17}
Alon Eden, Michal Feldman, Ophir Friedler, Inbal {Talgam{-}Cohen}, and
  S.~Matthew Weinberg.
\newblock A simple and approximately optimal mechanism for a buyer with
  complements.
\newblock In {\em Proceedings of the 2017 ACM Conference on Economics and
  Computation}, page 323, 2017.

\bibitem{Feige09}
Uriel Feige.
\newblock On maximizing welfare when utility functions are subadditive.
\newblock {\em SIAM Journal on Computing}, 39(1):122--142, 2009.

\bibitem{FFGL13}
Michal Feldman, Hu~Fu, Nick Gravin, and Brendan Lucier.
\newblock Simultaneous auctions are (almost) efficient.
\newblock In {\em Proceedings of the 45th Symposium on Theory of Computing},
  pages 201--210, 2013.

\bibitem{FGL15}
Michal Feldman, Nick Gravin, and Brendan Lucier.
\newblock Combinatorial auctions via posted prices.
\newblock In {\em Proceedings of the 26th Annual ACM-SIAM Symposium on Discrete
  Algorithms}, pages 123--135, 2015.

\bibitem{Friedman91}
Milton Friedman.
\newblock How to sell government securities.
\newblock {\em Wall Street Journal}, 1991.
\newblock August 28, 1991, page A8.

\bibitem{HKMN11}
Avinatan Hassidim, Haim Kaplan, Yishay Mansour, and Noam Nisan.
\newblock Non-price equilibria in markets of discrete goods.
\newblock In {\em Proceedings of the 12th {ACM} Conference on Electronic
  Commerce}, pages 295--296, 2011.

\bibitem{LLN06}
Benny Lehmann, Daniel~J. Lehmann, and Noam Nisan.
\newblock Combinatorial auctions with decreasing marginal utilities.
\newblock {\em Games and Economic Behavior}, 55(2):270--296, 2006.

\bibitem{LucierP11}
Brendan Lucier and Renato {Paes Leme}.
\newblock {GSP} auctions with correlated types.
\newblock In {\em Proceedings of the 12th {ACM} Conference on Electronic
  Commerce}, pages 71--80, 2011.

\bibitem{LucierPT12}
Brendan Lucier, Renato {Paes Leme}, and {\'{E}}va Tardos.
\newblock On revenue in the generalized second price auction.
\newblock In {\em Proceedings of the 21st World Wide Web Conference}, pages
  361--370, 2012.

\bibitem{MT15}
Evangelos Markakis and Orestis Telelis.
\newblock Uniform price auctions: Equilibria and efficiency.
\newblock {\em Theory Comput. Syst.}, 57(3):549--575, 2015.

\bibitem{Nisan15}
Noam Nisan.
\newblock Algorithmic mechanism design through the lens of multi-unit auctions.
\newblock In H.~Peyton Young and Shmuel Zamir, editors, {\em Handbook of Game
  Theory with Economic Applications, Volume 4}, chapter~9, pages 477--515.
  Elsevier, 2015.

\bibitem{PaeslemeT10}
Renato {Paes Leme} and {\'{E}}va Tardos.
\newblock Pure and {B}ayes-{N}ash price of anarchy for generalized second price
  auction.
\newblock In {\em Proceedings of the 51th Annual {IEEE} Symposium on
  Foundations of Computer Science}, pages 735--744, 2010.

\bibitem{Parkes07}
David~C. Parkes.
\newblock Online mechanisms.
\newblock In Noam Nisan, Tim Roughgarden, \'{E}va Tardos, and Vijay Vazirani,
  editors, {\em Algorithmic Game Theory}, chapter~16, pages 411--439. Cambridge
  University Press, 2007.

\bibitem{googleipo}
Jay Ritter.
\newblock Google's {IPO}, 10 years later.
\newblock
  \url{http://www.forbes.com/sites/jayritter/2014/08/07/googles-ipo-10-years-later},
  2014.
\newblock Accessed 2017-02-09.

\bibitem{Varian07}
Hal~R. Varian.
\newblock Position auctions.
\newblock {\em International Journal of Industrial Organization},
  25(6):1163--1178, 2007.

\end{thebibliography}

\end{document}